\documentclass[10pt,twocolumn,english,aps,pra,superscriptaddress]{revtex4-1}
\usepackage{color}
\usepackage{latexsym}
\usepackage{amsmath}
\usepackage{amssymb}
\usepackage{graphicx}
\renewcommand{\vec}[1]{{\bf #1}}
\newcommand{\mat}[1]{{\bf #1}}

\begin{document}

\title{
Supersymmetry ``protected'' topological phases of isostatic lattices and kagome antiferromagnets
}
\author{Michael J. Lawler}
\affiliation{Department of Physics, Binghamton University, Binghamton, NY, 13902, USA.}
\affiliation{Laboratory of Atomic And Solid State Physics, Cornell University, Ithaca, NY 14853, USA.}

\date{April 1$^\text{st}$, 2016}

\begin{abstract}
I generalize the theory of phonon topological band structures of isostatic lattices to frustrated antiferromagnets. I achieve this with a discovery of a many-body supersymmetry (SUSY) in the phonon problem of balls and springs and its connection to local constraints satisfied by ground states. The Witten index of the SUSY model demands the Maxwell-Calladine index of mechanical structures. ``Spontaneous supersymmetry breaking'' is identified as the need to gap all modes in the bulk to create the topological isostatic lattice state. Since ground states of magnetic systems also satisfy local constraint conditions (such as the vanishing of the total spin on a triangle) I identify a similar SUSY structure for many common models of antiferromagnets including the square, triangluar, kagome, pyrochlore nearest neighbor antiferromagnets, and the $J_2 = J_1/2$ square lattice antiferromagnet. Remarkably, the kagome family of antiferromagnets is the analog of topological isostatic lattices among this collections of models. Thus, a solid state realization of the theory of phonon topological band structure may be found in frustrated magnetic materials. 
\end{abstract}

\maketitle

Recently, Kane and Lubensky\cite{Kane2014} (KL) identified topological properties of isostatic lattice phonons. They achieved this by discovering a topological index governing mechanical structures by building on Calladine's work\cite{Calladine1978} and further utilizing a Dirac-like square-root of the phonon equations of motion, a problem to which they could apply the theory of topological insulators. Remarkably, they showed the existence of lattices with gapped phonons for periodic boundary conditions that must have gapless phonons with open boundary conditions. It is natural to wonder if this striking effect is more general and they conclude their study with ``Finally, it will be interesting to explore connections with theories of frustrated magnetism$^{48}$.'' where reference 48 is my study\cite{Lawler2013} identifying topological gauge dynamics of the zero modes of classical kagome antiferromagnets. This paper's goal is to make this connection and use it to generalize KL theory beyond phonons.

The KL theory of isostatic lattices is a different branch of topological phases from the theory of topological band insulators. Following the original discovery of topological insulators, topological properties of boson band structures have been studied for a wide variety of systems including phonons\cite{Prodan2009a,Peano2015}, photons\cite{Lu2014}, acoustic phononic crystals\cite{Xiao2015} and magnons\cite{Zhang2013,Mook2014}. These systems achieve their topological properties in the presence of time reversal symmetry breaking and are built directly from the physics of the integer quantum Hall effect. In contrast, KL's theory of isostatic phonons is time reversal symmetric and ``purely geometric in nature''\cite{Witten2015}. A connection to the integer quantum Hall effect is made only after a Dirac-like square-rooting procedure of the equations of motion. It therefore presents a new direction in the theory of topological phases. 

Remarkably, though its may not apply directly to solid state phonons because they are mechanically stable, the KL theory has already seen a variety of applications due to its insight into the general phenomena of mechanical collapse. These include some topological aspects of the jamming and rigidity percolation transitions\cite{Lubensky2015}, metamaterials made from beams and pins\cite{Chen2014}, and, remarkably,  origami\cite{Chen2016}. Ref. \onlinecite{Chen2014} has also taken it beyond the linearized level and discovered solitons that can propagate freely in the bulk of the isostatic state. So, given the fundamental insight it provides, any extension of KL theory to a new class of systems, including extensions that go beyond the linearized limit, is likely to shed new light on those systems. 

In this light, frustrated magnets and/or highly frustrated magnets are a prime target for an extension of KL theory. They are magnets not only ``on the verge of collapse'' but also those that have already ``collapsed''. Here collapsing is the analog of destabilizing the magnetic ordered state into a paramagnetic state such as a quantum spin liquid or valence bond solid.  A variety of materials including the organics, kagome family and pyrochlores are heavily studied for this reason\cite{Balents2010}.  In addition, highly geometrically frustrated magnets have a form of accidental degeneracy that results from a special feature of the spin Hamiltoinan\cite{Moessner2001}. This frustration is toy-like (fine tuned) perhaps in a similar way that balls and springs are toy-like versions of a general theory of phonons. So if KL theory were applicable to frustrated magnets, this generalization might apply to many already realized solid state materials.

In this paper, we show that the key to generalizing KL theory to other systems lies in connecting it to an abstract theoretical framework that takes the form of a many-body supersymmetric (SUSY) structure that extends the description of balls and springs. The new fermionic degrees of freedom, that I dub ``phoninos'', are superpartners to phonons and are governed by the KL theory's square-rooted equations of motion. For linearized phonons, the two sets of degrees of freedom are decoupled. The phoninos therefore need not be real degrees of freedom but just reflect the specialness of balls and springs compared to a more general theory of phonons. I then show that the topological index identified in KL theory is demanded by the Witten index\cite{Witten1982} of SUSY. A similar conclusion was reached by Ref. \onlinecite{Vitelli2014} in the continuum limit. Remarkably, spontaneous SUSY breaking here is just the need to gap all modes in the bulk to create the topological state (as could be the case for topological superconductors\cite{Grover2014}). Finally, we discuss what protects this topology including the role of quantum effects and non-linearities. 

I then apply the same many-body SUSY construction to the case of magnons. It turns out SUSY is found for both unfrustrated and frustrated magnons for either case has ground states that satisfy local constraints and the SUSY construction is built on these constraints. Remarkably, kagome magninos are governed by a Hamiltonian which is the Dirac's constraint matrix\cite{Dirac1950,Dirac1958} studied in Ref. \onlinecite{Lawler2013} and is related through SUSY to the ordinary kagome magnon problem. The Witten index in this case is then shown to vanish for periodic boundary conditions but not for open boundary conditions demonstrating that kagome magnons are the analog of the isostatic lattice of KL theory if we can gap all their modes without changing topology such as in a distorted kagome crystal. Such a phase we call isostatic magnetism. We conclude by identifying some of the new physical phenomena predicted by these formal development: the design of magnets by exploiting weaknesses in isostatic magnetism such as through a rich set of magnetic field induced spin flip transitions that we call ``self field modes''; and an application of the ideas here to the origin of the industrial strength of Mn$_3$Ir (why it is a robust antiferromagnet); the potential application of topological band theory of electrons to magninos in the isostatic magnetic phase and via SUSY to magnons; the study of the loss of magnetic order via isostatic magnetism which can be viewed as a classical critical point between magnetic order and a classical spin liquid\cite{Moessner1998a}/cooperative paramagnet\cite{Villain1979}.

\section{Topological aspects of phonons}
\subsection{Theoretical constructs in Phonon Models}\label{sec:PhononModel}
To generalize KL theory to magnetism, we need to first identify the theoretical constructs that underpin the theory of phonons and then seek analogs of these constructs in the magnon problem. 

The simplest description of phonons, that of the vibrations of balls connected to springs (in the classical limit), is endowed with a number of theoretical constructs that shed much light on their behavior. To begin with, this simplicity demands a specific form of their Hamiltonian. If we define the extension of spring $m$ to be $e_m$, then the most general form for an ideal balls and springs classical Hamiltonian is
\begin{equation}\label{eq:Hphonon}
H_{phonon} = \frac{1}{2}p_{i\alpha}m^{i\alpha,j\beta}p_{j\beta} +  \frac{1}{2}e_mk^{mn} e_{n}
\end{equation}
Here $p_{i\alpha}$ is the $\alpha=\{x,y,\ldots\}$ component of the momentum of the ball labeled by $i$, $m^{i\alpha,j\beta}$ is the matrix inverse of the ``mass tensor'' $m_{i\alpha,j\beta}$, $k^{mn}$ is the spring constant matrix and repeated indices are summed over. In the simplest setting, $m_{i\alpha,j\beta} = m\delta_{ij}\delta_{\alpha\beta}$ and $k^{mn} = k\delta^{mn}$ are proportional to identity matrices. Here we leave them in the general form to aid our study of the structure of the theory and not its application. The restriction of the Hamiltonian to that of balls and springs therefore introduces two (inverse) metrics $k^{mn}$ and $m^{i\alpha,j\beta}$ in configuration space and momentum space respectively and identifies the ground state as satisfying constraints $p_{i\alpha}=0$, $e_m=0$. 

Balls and springs are a toy-like simplification of a full phonon problem. We will return to the justification of their use and the validity of the following results in the next section (section \ref{sec:topologicalprotections}). 

In the linearized phonon limit, there is also a matrix $A_{m,i\alpha}$ and an associated topological invariant. This matrix relates $e_m$ to the displacements of each atom $u^{i\alpha}$ from their equilibrium positions via $e_m = A_{m,i\alpha} u^{i\alpha}$. This matrix also relates the forces $F^{i\alpha}$ to the tensions $T^m$ in each spring via $F^{i\alpha} = A^T_{i\alpha,m} T^m$. The topological invariant associated with this matrix (which we will call the Maxwell-Calladine index\cite{Calladine1978} though it was discovered by Kane and Lubensky) is
\begin{align}
   \nu &= dN_s - N_b \\
         & = (\text{rank} \mat A + \text{nullity} \mat A) - (\text{rank} \mat A^T + \text{nullity} \mat A^T)\\
         & = \text{nullity} \mat A - \text{nullity} \mat A^T
\end{align} 
where $d$ is the number of components $\alpha$, $N_s$ the number of sites, $N_b$ the number of springs (bonds), $\text{nullity}\, \mat{M}$ denotes the dimension of the null space of $\mat{M}$ and we have used $\text{rank}\, \mat A = \text{rank}\, \mat A^T$ by the fundamental theorem of linear algebra. This quantity relates the topology data of the system, $d$, $N_s$ and $N_b$ to the number of zero modes $N_0=\text{nullity} \mat A$ and number of states of self stress $N_{ss} = \text{nullity}\mat A^T$. Here a zero mode is a vector $u^{i\alpha}_0$ that satisfies $e_m = A_{m,i\alpha}u^{i\alpha}_0 = 0$ and a state of self stress is a vector of tensions $T^m_0$ that satisfies $F^{i\alpha} = A^T_{i\alpha,m}T^m_0 = 0$. This quantity is topological in the sense that it doesn't depend on the metrics $k^{m,n}$ and $m^{i\alpha,j\beta}$ and remains unchanged under any changes of the matrix $A_{m,i\alpha}$. So linearized balls and springs also have a natural topological invariant.

Kane and Lubensky argued that a Dirac-like square-rooted equations of motion enables the study of phonons following the theory of topological insulators. These equations of motion are defined by the matrix
\begin{equation}
  {\mathcal H} = \begin{pmatrix} 0 & -m^{i\alpha,k\gamma}A^T_{k\gamma,n}\\ k^{m,p} A_{p,j\beta} & 0\end{pmatrix}
\end{equation}
Here I have taken the liberty to add a minus sign to Kane and Lubensky's matrix and re-inserted $k^{m,n}$ and $m^{i\alpha,j\beta}$ which they set to identity matrices. This is so that squaring this matrix produces
\begin{equation}
  {\mathcal H}^2 = \begin{pmatrix} -m^{i\alpha,k\gamma}A^T_{k\gamma,p}k^{p,q} A_{q,j\beta} & 0\\ 0 & -k^{m,p} A_{p,k\gamma}m^{k\gamma,l\delta}A^T_{l\delta,n}\end{pmatrix}
\end{equation}
where the upper left block defines the second order differential equation for the displacements 
\begin{equation}
   \ddot u^{i\alpha} = -m^{i\alpha,k\gamma}A^T_{k\gamma,p}k^{p,q} A_{q,j\beta} u^{j\beta}
\end{equation}
The minus sign can be removed by multiplying $\mathcal{H}$ by $\tau^z = \left(\begin{smallmatrix} I& 0\\0&-I\end{smallmatrix}\right)$ from the right, something we are free to do since $\tau^z$ commutes with the ${\mathcal{H}}$ (a feature noted by Kane and Lubensky). 

This model helps us understand the importance of the topological index. If, for periodic boundary conditions, $\nu\neq 0$ then there must always be either a zero mode (if $\nu>0$) or a self stress mode (if $\nu<0$). In the spectrum of $\mathcal{H}$ there is no gap to the excitations. However, if $\nu=0$ then we can always find an $A_{m,i\alpha}$ such that the gap disappears. In such a case, if we open the boundary conditions we will invariably find that $\nu\neq 0$ since we change the counting of the number of sites and bonds and so either special self stresses exist due to boundary conditions or special zero modes exist. In either case, there is an interplay between a gap, topology and boundary conditions as found in other topological systems such as the quantum Hall effect or topological insulators.

Finally, there is one more mathematical object we will find useful for the study of phonons. To study the zero modes of this problem directly, we can follow Ref. \onlinecite{Lawler2013} and view the Hamiltonian as energetically imposing constraints on the degrees of freedom. Here these constraints are simply $p_{i\alpha} = 0$ and $e_m=0$ as noted above for if these conditions are met the Hamiltonian vanishes. In his development of constrained Hamiltonian mechanics, Dirac pointed out that an important object in the study of constraints in phase space is the constraint matrix which here takes the form:
\begin{equation}\label{eq:PhononCmatrix}
{\mat C} = \begin{pmatrix} \{ p_{i\alpha},p_{j\beta}\} & \{ p_{i\alpha},e_n\} \\ \{ e_m,p_{j\beta}\} & \{e_m,e_n\} \end{pmatrix} 
\end{equation}
where $\{\cdot,\cdot\}$ is the usual Poisson bracket of classical mechanics. A vector in the null space of this matrix is then either associated with a redundant constraint (when there are more constraint functions than necessary to constrain the variables) or with a zero mode coordinate that has no conjugate variable in the space of zero modes (gauge coordinate). If we expand the spring extensions $e_m$ in terms of the displacements of the atoms from their equilibrium positions $u^{i\alpha}$ as $e_m = A_{m,i\alpha}u^{i\alpha}$ then the constraint matrix takes the form
\begin{equation}
{\mat C} = \begin{pmatrix} 0 & -A^T_{i\alpha,n} \\ A_{m,j\beta} & 0 \end{pmatrix} 
\end{equation} 
Remarkably, in this form, $\mat{C}$ is similar to the ``square-rooted'' Hamiltonian $\mathcal{H}$ discussed above. If we work in Kane and Lubensky's units where $k^{m,n} = \delta^{m,n}$ and $m^{i\alpha,j\beta} = \delta^{ij}\delta^{\alpha\beta}$ they are actually the same. We also see that the redundant constraints are associated with self stress modes living in the null space of $A^T_{i\alpha,m}$ and the gauge coordinates with the zero modes living in the null space of $A_{m,i\alpha}$.  So the eigenvalue problem associated with Dirac's constraint matrix here is closely related to the Dirac-like square-rooted phonon Hamiltonian!

Now the relationship between all of these mathematical objects, the metrics, the Hamiltonians, the $A_{m,i\alpha}$ matrix, and the constraint matrix are not so easily understood. They all clearly derive from the balls and springs construction. But how is the topological index dependent on the metrics? How is the constraint matrix $C$ related to the square-rooted Hamiltonian $\mathcal{H}$?  In what way is the topology protected? Can these constructions be applied to other systems that are not balls and springs?

To address these questions, we need an abstract theoretical framework that identifies all of these mathematical object from a simpler set of constructs. Remarkably, such a framework is provided by extending the model's phase space to include a set of fermion degrees of freedom $\gamma^m$ and $\Gamma^{i\alpha}$ that we will call phonino modes and imposing supersymmetry. From this vantage point we will then look down on all the mathematical objects discussed previously and see their relationships to each other. The formal nature of the supersymmetry will then also allow us to connect two seemingly different problems: balls and springs and quadratic spin models. We then just need to determine if the connection is merely formal or whether it predicts new physical phenomena for either balls and springs or magnetic systems.

\subsection{Supersymmetric phonons}\label{sec:SUSYPhonons}
A Hamiltonian that is a quadratic form in constraint functions that define its ground state has a natural supersymmetry. To construct this supersymmetric model, we begin by defining the supersymmetric charge by the product of a constraint function and a new degree of freedom:
\begin{equation}
  Q = \Gamma^{i\alpha} p_{i\alpha} + \gamma^{m}e_{m}.
\end{equation}
This looks like a Lagrange multiplier term in an action meant to impose the constraints $p_{i\alpha} = 0$ and $e_m=0$ where $\Gamma^{i\alpha}$ and $\gamma^m$ are the Lagrange multipliers. However, here we view these as degrees of freedom that satisfy their own Poisson bracket relations. Indeed we can construct the supersymmetric Hamiltonian using the Poisson bracket
\begin{multline}
  H_{SUSY} = \frac{1}{2}\{Q,Q\} = \\
\gamma^m\{e_{m},p_{i\alpha}\} \Gamma^{i\alpha}+ \frac{1}{2}p_{i\alpha}\{\Gamma^{i\alpha},\Gamma^{j\beta}\}p_{j\beta} + \frac{1}{2}e_m\{\gamma^m,\gamma^n\}e_n
\end{multline}
where $\{f,g\}$ denotes a Poisson bracket. For this equation to be true, it is necessary for $\gamma^m$ and $\Gamma^{i\alpha}$ to be Grassmann numbers so that $Q$ is Grassmann odd and $\{Q,Q\}$ does not vanish by the normal antisymmetry of Poisson brackets. We also used $\{\Gamma^{i\alpha},\gamma^m\}=\{p_{i\alpha},p_{j\beta}\} = \{e_m,e_n\}=0$. If we choose the remaining Poisson brackets to be 
\begin{equation}
\{\Gamma^{i\alpha},\Gamma^{j\beta}\}=m^{i\alpha,j\beta},\quad \{\gamma^m,\gamma^n\}=k^{mn}
\end{equation}
then we obtain the simple relation $H_{SUSY} = H_{phonino} + H_{phonon}$ where $H_{phonino}$ is the first term and the second and third terms make up $H_{phonon}$ of Eq. \ref{eq:Hphonon}. This defines the Poisson bracket to be\cite{Casalbuoni1976b,Berezin1977,Nakano1980}
\begin{multline}
  \{f,g\} = \frac{\partial f}{\partial u^{i\alpha}}\frac{\partial g}{\partial p_{i\alpha}} - \frac{\partial f}{\partial p_{i\alpha}}\frac{\partial g}{\partial u^{i\alpha}} +\\
f\frac{\overleftarrow\partial}{\partial \Gamma^{i\alpha}}m^{i\alpha,j\beta}\frac{\partial g}{\partial \Gamma^{j\beta}} +
f\frac{\overleftarrow\partial}{\partial \gamma^m}k^{mn}\frac{\partial g}{\partial \gamma^n}
\end{multline}
which is symmetric if $f$ and $g$ are both Grassmann odd and antisymmetric otherwise. Finally, we see that $\{Q,H_{SUSY}\}=0$ so the two observables form a closed superalgebra. So we can promote the phonon problem to a supersymmetric problem of bosonic phonons and fermionic phoninos where the phoninos satisfy a Clifford algebra with metrics $k^{mn}$ and $m^{i\alpha,j\beta}$ and the phonino Hamiltonian is determined by the constraint matrix
\begin{equation}
  H_{phonino} = \frac{1}{2} \begin{pmatrix} \Gamma^{i\alpha} & \gamma_m\end{pmatrix}\mat{C}
  \begin{pmatrix}\Gamma^{j\beta}\\\gamma^n\end{pmatrix}.
\end{equation}
where $\mat{C}$ is given by Eq. \ref{eq:PhononCmatrix}. 

Note, here we have chosen to study phonons classically using Grassmann numbers. In this sense we follow one of the original motivations to introduce Grassmann numbers into classical mechanics: to construct simpler-than-field-theory supersymmetry models\cite{Casalbuoni1976b}. The use of classical mechanics here, however, is just to connect with the history of theories of balls and springs and is not necessary as we will discuss section \ref{sec:topologicalprotection}. Furthermore, our analysis of the model presented below will be informed by the use of supersymmetry in quantum mechanics (see Ref. \onlinecite{Cooper1994} for a review), because it is another simpler-than-field-theory supersymmetry model. However, unlike quantum mechanics, the above model is a many-body supersymmetric lattice model, one not so different from the lattice model of a supersymmetric topological superconductor studied in Ref. \onlinecite{Grover2014}.

In the linearized limit the two Hamiltonians become
\begin{equation}
  H_{phonon} = \frac{1}{2}p_{i\alpha}m^{i\alpha,j\beta}p_{j\beta} + \frac{1}{2}u^{i\alpha}A^T_{i\alpha,m}k^{mm'} A_{m',j\beta}u^{j\beta}
\end{equation} 
and
\begin{equation}
 H_{phonino} =\gamma^m A_{m,i\alpha}\Gamma^{i\alpha}
\end{equation}
where we used $\{p_{i\alpha},u^{j\beta}\} = \delta_i^j\delta^\alpha_\beta$. In this limit then, the phonons and the phoninos are decoupled and the Hamiltonians are quadratic.  We therefore have a model we can apply to real systems since it contains the correct phonon eigenvalue problem. It is not so  obvious, however, what we can learn by introducing the phoninos. To settle this, we need to directly study the different eigenvalue problems.

\subsection{The SUSY eigenvalue problem} \label{sec:PhononEigenmodes}
To study the eigenvalue problems of $H_{SUSY}$, we need to solve the corresponding equations of motion. The first order in time differential equations for the phonons are
\begin{equation}
  \begin{pmatrix} \dot u^{i\alpha}\\ \dot p_{i\alpha}\end{pmatrix} = 
      \begin{pmatrix} 0 & m^{i\alpha,j\beta}\\ -A^T_{i\alpha,m}k^{m,n}A_{n,j\beta}\end{pmatrix}
  \begin{pmatrix} u^{j\beta}\\ p_{j\beta}\end{pmatrix}      
\end{equation}
and for the phoninos are
\begin{equation}\label{eq:SqrtedEOM}
  \begin{pmatrix} \dot \Gamma^{i\alpha}\\ \dot \gamma^m\end{pmatrix} = 
      \begin{pmatrix} 0 & -m^{i\alpha,k\gamma}A^T_{k\gamma,n} \\ k^{m,p}A_{p,j\beta}& 0\end{pmatrix}
  \begin{pmatrix} \Gamma^{j\beta}\\ \gamma^{n}\end{pmatrix}      
\end{equation}
Remarkably, the phonino equations of motion are the square-rooted hamiltonian matrix $\mathcal{H}$ introduced by Kane and Lubensky. The constraint matrix $\mat{C}$ and $\mathcal{H}$ are therefore related in the supersymmetric model by one entering the Hamiltonian quadratic form and the other by the corresponding equations of motion. 

We will also find it useful to work with the second derivative in time equations of motion:
\begin{eqnarray}
 \ddot u^{i\alpha} &=& -m^{i\alpha,j\beta}A^T_{j\beta,m}k^{m,n}A_{n,k\gamma}u^{k\gamma}\\
 \ddot p_{i\alpha} &=& -A^T_{i\alpha,m}k^{m,n}A_{n,j\beta}m^{j\beta,k\gamma}p_{k\gamma}\\
 \ddot \Gamma^{i\alpha} &=& -m^{i\alpha,j\beta}A^T_{j\beta,m}k^{m,n}A_{n,k\gamma}\Gamma^{k\gamma}\\
 \ddot \gamma^{m} &=& -k^{m,n}A_{n,i\alpha}m^{i\alpha,j\beta}A^T_{j\beta,p}\gamma^{p}
\end{eqnarray}
From these, we see that, $u^{i\alpha}$, $v^{i\alpha}=m^{i\alpha,j\beta}p_{j\beta}$ and $\Gamma^{i\alpha}$ all obey the same equations of motion. $\gamma^m$ obeys a distinctly different equation. We also can determine that each equation has two types of eigenmodes. For the $u^{i\alpha}$-type equations, all modes either have zero eigenvalue and are in the null space of $A_{m,i\alpha}$ or they have finite eigenvalues. There are no modes with zero eigenvalue that satisfy $A_{m,i\alpha}u^{i\alpha}\neq0$. This is because such a mode would satisfy $g_mk^{m,n}g_n=0$ with $g_m=A_{m,i\alpha}u^{i\alpha}\neq 0$ and violate the assumption that $k^{m,n}$ is a positive definite matrix. For a similar reason, the $\gamma^m$ equation has two modes types, those with zero eigenvalue in the nullspace of $A^T_{i\alpha,m}$ and those with finite eigenvalues. 

We can of course proceed by solving these equations separately, but this would not give us any insight into how the phoninos can help us understand the phonons. Instead, lets proceed by studying supersymmetry transformations and reduce the problem to solving only the parts not related by supersymmetry.

The supersymmetric charge provides us with a map between the phase space observables with an even number of phonino coordinates (Grassmann even) and those with an odd number (Grassmann odd). The single particle phonon modes which are linear combinations of $u^{i\alpha}$ and $p_{i\alpha}$ are in the even group while the single particle phonino modes which are linear combinations of $\Gamma^{i\alpha}$ and $\gamma^m$ are in the odd group. The map $\{\cdot,Q\}$ we can construct from $Q$ sends an observable from the even sector to the odd sector or vice versa by inserting it into the location of the $\cdot$. This map has a special relationship with the equations of motion: time evolution of the mapped observable is the same as the unmapped observable. For example, $\{u^{i\alpha},Q\}$ is a phonino observable that obeys the equations of motion
\begin{equation}
   \frac{d^2}{dt^2} \{u^{i\alpha},Q\} = -m^{i\alpha,j\beta}A^T_{j\beta,m}k^{m,n}A_{n,k\gamma} \{u_{k\gamma},Q\}
\end{equation}
which is the equations of motion for $u^{i\alpha}$. Similarly, the phonon observables $\{p_{i\alpha},Q\}$, $\{\Gamma^{i\alpha},Q\}$, $\{\gamma^m,Q\}$ obey the $p_{i\alpha}$, $\Gamma^{i\alpha}$ and $\gamma^m$ equations respectively. Computing these Poisson brackets explicitly, we see that they are
\begin{eqnarray}
\{u^{i\alpha},Q\} = \Gamma^{i\alpha},\quad \{p_{i\alpha},Q\} = A^T_{i\alpha,m}\gamma^m\\
\{\Gamma^{i\alpha},Q\} = m^{i\alpha,j\beta}p_{j\beta},\quad \{\gamma^m,Q\} = k^{m,n}A_{n,i\alpha}u^{i\alpha}
\end{eqnarray}
So it would seem that supersymmetry is too powerful a symmetry, that the phonino problem is equivalent to the phonon problem (and therefore uninteresting). 

To study these relationships in more detail, lets now take linear combinations of phase space observables. If $v^{i\alpha}p_{i\alpha}$ is an eigenmode of the $p_{i\alpha}$ equation then it obeys
\begin{equation}
  v^{i\alpha} A^T_{i\alpha,m}k^{m,n}A_{n,j\beta}m^{j\beta,k\gamma} = \omega^2 v^{k\gamma}
\end{equation} 
Now if $v^{i\alpha}A^T_{i\alpha,m} = 0$ so the $v^{i\alpha}$ is in the (left) nullspace of $A^T_{i\alpha,m}$, then it is an eigenvector with frequency $\omega=0$. If this is not the case, it is an eigenvector with a finite frequency as discussed above. Now lets map this to a linear combination of $\gamma^m$'s using $Q$:
\begin{equation}
  \{v^{i\alpha}p_{i\alpha},Q\} = v^{i\alpha} A^T_{i\alpha,m}\gamma^m
\end{equation}
If $v^{i\alpha}$ therefore corresponds to a zero mode, it gets annihilated by this map. Only the finite frequency modes therefore actually pass from the $p_{i\alpha}$ eigenmode problem to the $\gamma^m$ eigenmode problem. However, for the finite mode case, notice that the mapped observable is a linear combination whose coefficients obey
\begin{equation}
  v^{i\alpha}A^T_{i\alpha,m}k^{m,n}A_{n,j\beta}m^{j\beta,k\gamma}A^T_{k\gamma,p} = \omega^2 v^{i\alpha}A^T_{i\alpha,p}
\end{equation}
So it is a (left) eigenvector of the $\gamma^m$ differential equation with the same eigenvalue it had in the $p_{i\alpha}$ differential equation before it was mapped. Mapped observables therefore carry their eigenvalue with them.

\begin{figure}
\includegraphics[width=0.6\columnwidth]{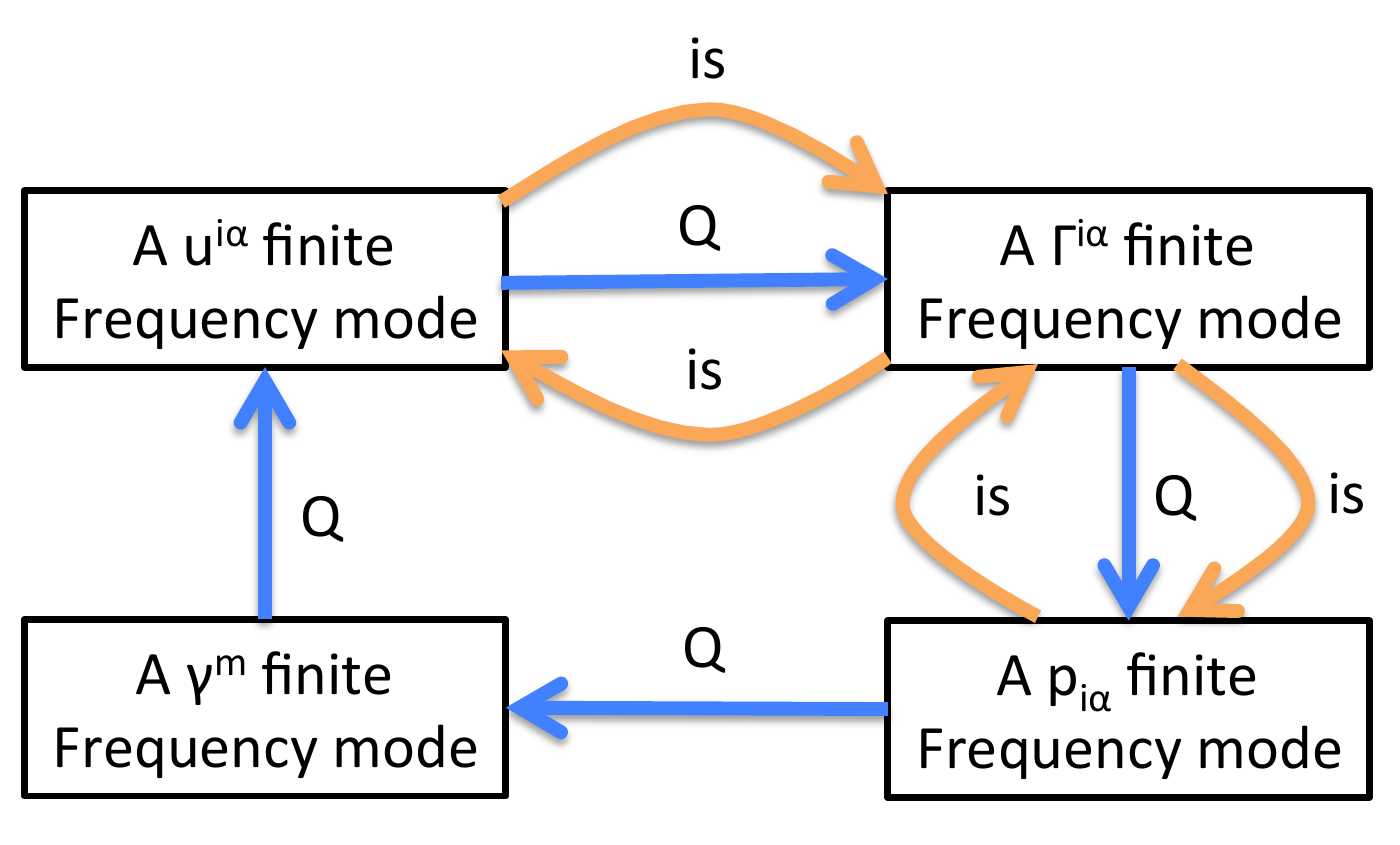}
\includegraphics[width=0.6\columnwidth]{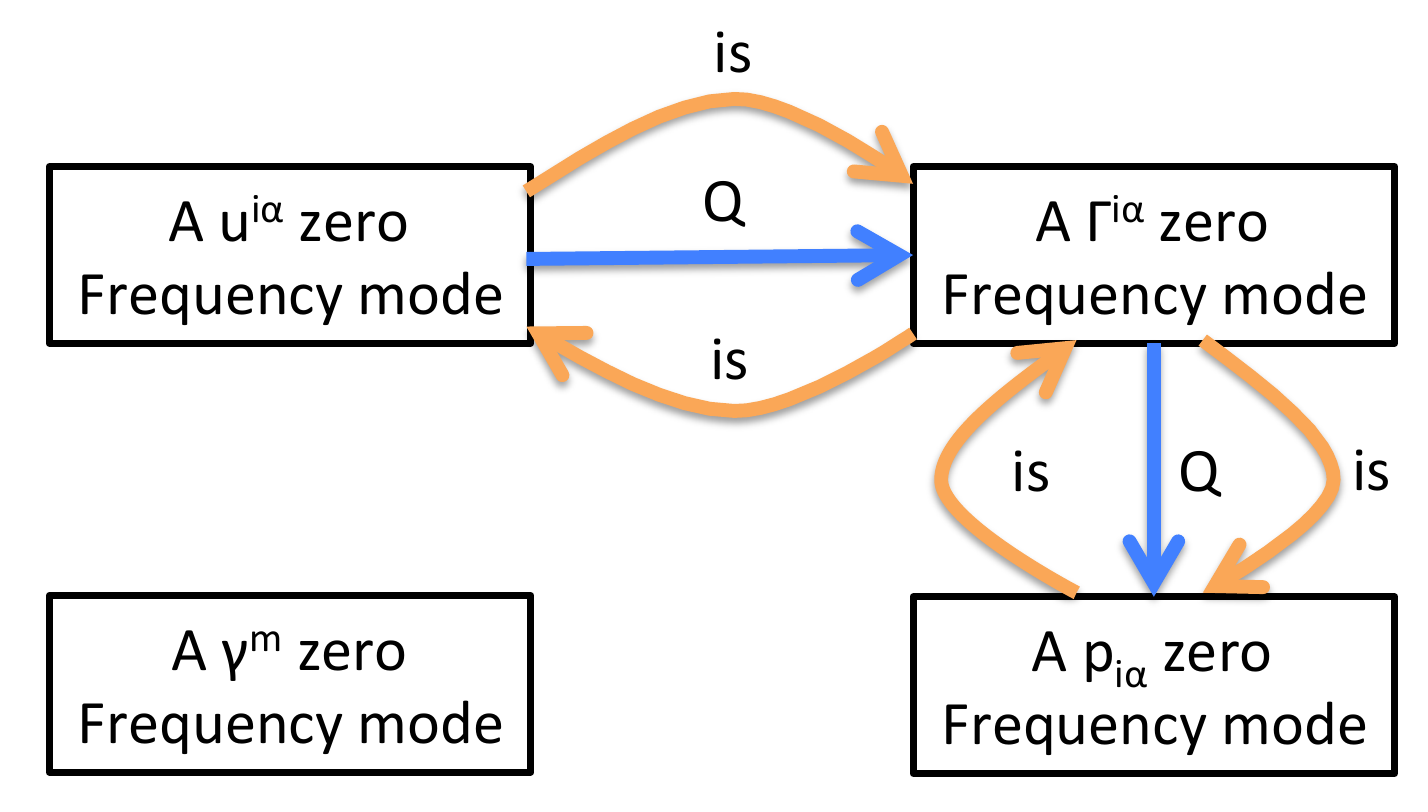}
\caption{Graphical display of eigenmode relationships. Here two types of arrows are drawn: one called ``is'' results from the identical form of the two differential equations and one called ``$Q$'' is the supersymmetry map. Notice $Q$ does not map zero mode eigenstates from $p_{i\alpha}$ to $\gamma^m$ and from $\gamma^m$ to $u^{i\alpha}$. Since all arrows represent maps of unique elements to unique elements, if a path of arrows exists from one set of modes to another and another path exists from that set of modes back to the first then the two sets of modes are isomorphic. For finite modes, any two sets of eigenstates have such paths relating them so they are all isomorphic to each other. For zero modes, only the $u^{i\alpha}$, $\Gamma^{i\alpha}$ and $p_{i\alpha}$ sets of eigenstates are isomorphic. The $\gamma^m$ set is different and hence the phonino problem has a different set of zero modes from the phonon problem. Note this figure follows the rules of an olog\cite{Spivak2014}.}\label{fig:PhononQMap}
\end{figure}

Extending this study to each of the four differential equations and breaking their set of eigenvectors into two groups, the finite and the zero eigenvalue modes, we have constructed Fig. \ref{fig:PhononQMap} that graphically depicts all the different relationships. In it an arrow represents a function mapping the set of eigenmodes at its tail to the set of eigenmodes at its head. If two different paths exist between any two sets, one going from one set to the other and the other going the reverse direction, then the two sets are isomorphic and in particular have the same number of eigenmodes. In this way, we see that the number of finite energy eigenmodes are the same for all differential equations, just their zero modes are different.

Now, with the above analysis of the eigenmodes, we are in a position to understand the topological index associated with a supersymmetric theory. As pointed out by Witten (and often called the ``Witten index'') the total number of boson modes minus the total number of fermion modes ($Tr (-1)^F$ where $F=0$ for a boson mode, and $F=1$ for a fermion mode) is a topological invariant. Applying this to the single particle sector we have
\begin{equation}
  \nu = Tr(-1)^F = 2dN_s - dN_s - N_b = dN_s - N_b
\end{equation}
But if we break down the count of zero modes into the four groups of two types of modes, we have
\begin{align}
  \nu &= \underbrace{(N_{>} + N_0)}_\text{$u^{i\alpha}$ modes} + \underbrace{(N_{>} + N_0)}_\text{$p_{i\alpha}$ modes} - 
\underbrace{(N_{>} + N_0)}_\text{$\Gamma^{i\alpha}$ modes} - \underbrace{(N_> + N_{ss})}_\text{$\gamma^m$ modes}\\
        & = N_0 - N_{ss}
\end{align}
where we used Fig. \ref{fig:PhononQMap} to establish that the number of finite modes $N_{>}$ is the same between all modes types while $N_0$ is the same only between $u^{i\alpha}$, $\Gamma^{i\alpha}$ and $p_{i\alpha}$ and is in general different from the number of $\gamma^m$ zero modes, $N_{ss}$. To derive these results, we heavily relied on the positive definiteness of $m^{i\alpha,j\beta}$ and $k^{m,n}$ for if this were not the case, more modes could fail to be mapped by $Q$. So the Witten index of supersymmetry, when applied to the single particle sector valid for the linearized theory, is identical to the Maxwell-Calladine index but its derivation is directly built on eigenvalue problems rather than directly on the properties of the matrix $A_{m,i\alpha}$. 

The Witten index is useful in a supersymmetric theory because it dictates whether there must be a zero mode. If $\nu > 0$ there must be $\nu$ bosonic (phonon) zero modes. If $\nu < 0$ there must be $|\nu|$ fermionic (phonino) zero modes which here must be self stress modes. So long as $\nu\neq0$ then, supersymmetry can exist in the ``ground state'' for there is then a zero mode that satisfies $\hat Q|0> = 0$ (in the quantum language). However, if $\nu=0$ then without explicitly breaking supersymmetry one can remove all zero modes. Then there are no states satisfying $\hat Q|0> = 0$ and supersymmetry is spontaneously broken in the ground state. Remarkably, an isostatic lattice has both $\nu=0$ for periodic boundary conditions and all zero modes gapped ($N_0=0$). So isostatic lattices are an example spontaneous SUSY breaking.

Finally, we need to comment on the question of how the above supersymmetry is related to the quantum mechanics-like supersymmetry of the square rooted equations of motion (Eq. \ref{eq:SqrtedEOM}) pointed out by Kane and Lubensky. This symmetry is a supersymmetry between $\Gamma^{i\alpha}$ modes and $\gamma^m$ modes (both of which are fermionic here) and has an associated topological index which is numerically the same index as the single particle Witten index discussed above and the Maxwell-Calladine index. So the phonon problem appears to have two different supersymmetries, one at the full phase space level that we discuss above and the other at the level of the equations of motion of the linearized fermions that related different fermion modes to each other (quantum mechanics-like SUSY need not always relate bosonic modes to fermionic modes). 

In summary, the supersymmetry model has brought all mathematical objects mentioned so far under one umbrella. The supersymmetric theory has the phonon Hamiltoinan, the two metrics $m^{i\alpha,j\beta}$ and $k^{m,n}$ play a central role as they determine the commutation relations of the fermions, the phonino Hamiltonian is determined by the constraint matrix $\mat{C}$ and its corresponding equations of motion is the square-rooted Hamiltonian $\mathcal{H}$. Finally, the topological index is determined by the supersymmetry relations between the various eigenmodes. Lastly, it suggests that the specialness of the balls and springs description is the requirement that it is supersymmetric. If we break this specialness, if we explicitly break the supersymmetry, then most of the mathematical objects contained in the theory loose their meaning. In this way, I argue that the topological structure of isostatic lattices identified by Kane and Lubensky is SUSY protected.

\section{Topological Protection} \label{sec:topologicalprotection}
The previous results show that models of phonons built on constraint functions like the extension $e_m$ of a spring from its equilibrium length have a topological index tied to their equations of motion by a supersymmetric relationship with phoninos. This index guarantees the existence of zero modes from topology (even in the absence of a continuous symmetry) much like Goldstone's theorem guarantees the existence of zero modes from spontaneous symmetry breaking. Here we discuss the question of to what extent this topological property is robust.

\subsection{Robustness to quantum fluctuations}
For the most part, we worked in the classical limit where phonons are actually lattice vibrations. If we quantize it, then we convert Poisson brackets to commutators or anticommutators such as $[\hat p_{i\alpha},\hat u^{j\beta}] = i\delta_{i\alpha}^{j\beta}$ and $\{\hat\gamma^{m},\hat\gamma^{n}\}= k^{m,n}$. Quantizing the supersymmetric charge $\hat Q = \hat\Gamma^{i\alpha}\hat p_{i\alpha}+ \hat\gamma^{m}A_{m,i\alpha}\hat u^{i\alpha}$ then gives the supersymmetric Hamiltonian
\begin{multline}
\hat H_{SUSY} = \frac{1}{2}\{\hat Q,\hat Q\} = i\hat\gamma^m A_{m,i\alpha}\hat\Gamma^{i\alpha} +\\ 
 \frac{1}{2}\hat p_{i\alpha}m^{i\alpha,j\beta}\hat p_{j\beta} + \frac{1}{2}\hat u^{i\alpha}A^T_{i\alpha,m}k^{mm'} A_{m',j\beta}\hat u^{j\beta}
\end{multline}
This is identical in form to the classical case. A similar correspondence between the non-linear forms of the classical and quantum model also holds.  So the main results of this paper hold in both the classical and quantum forms and are not restricted to semi-classical or classical regimes.

\subsection{What do metric distortions do?}
Since the metrics $k^{m,n}$ and $m^{i\alpha,j\beta}$ entering the phonon problem are not present in the respective topological indices, the topological protection holds if these undergo a distortion so long as they remain positive definite matrices. In particular, the topological index holds if we stretch and distort the lattice as was used extensively in the discovery of Weyl phonon modes in Ref. \onlinecite{Rocklin2016}.

\subsection{What happens in the non-linear phonon problem?}
Let us first address this question for the non-linear supersymmetry models before returning to the problem of purely non-linear phonons in the absence of phoninos. 

In the presence of non-linearities, the eigenmodes of the system are no longer linear combinations of single particle observables. Instead, they become a linear combination of all multiparticle observables. As a result, the Witten index restricted to the single particle sector looses its meaning but the full index
\begin{align}
  \nu &= Tr(-1)^F\\ &= \text{\# of bosonic observables} - \text{\# of fermionic observables}
\end{align} 
still holds because $Q$ still maps multiparticle bosonic observables to a multiparticle fermionic observables. To get a sence of this full set of observables, we have considered the two-particle sector. This consists of bosonic observables of types $u^{i\alpha}u^{j\beta}$, $p_{i\alpha}p_{j\beta}$, $u^{i\alpha}p_{j\beta}$, $\gamma^m\gamma^n$, $\Gamma^{i\alpha}\Gamma^{j\beta}$ and $\gamma^m\Gamma^{i\alpha}$ and fermionic observables of types $\gamma^mu^{i\alpha}$, $\gamma^mp_{i\alpha}$, $\Gamma^{i\alpha}u^{j\beta}$ and $\Gamma^{i\alpha}p_{j\beta}$. Counting all of these two different two particle observables, noting that squares of grassmann observables vanish (i.e. $\gamma^m\gamma^m=0$) I obtain after simplifying
\begin{equation}
   \nu^{\text{2 particle}} = dN_s(dN_s+1)/2 + N_b(N_b-1)/2 -  dN_sN_b
\end{equation}
If $N_b=dN_s$ as in the isostatic lattice case, we see this also vanishes. Since the index for the one-particle and two-particle sectors vanishes, likely the total sum over all sectors relevant for the many body non-linear problem also vanishes. So the topological protection remains in effect for the full non-linear supersymmetric problem.

With this technical extension, we see that the non-linear models are also topological. Since they consist now of majorana fermions coupled to phonons, they could in principle be realized in a carefully engineered superconductor. In practice, however, they may prove to be more useful as a means of gaining theoretical insight into ``symmetry protected topological order'' in the presence of interacting fermions which does not make use of K-theory\cite{Kiteav2009}, Chern-Simons quantum field theory\cite{Levin2012,Lu2012} or supercohomology\cite{Gu2014}. 

Unfortunately, the salient feature of the non-linear problem is that the phonons are now coupled to the phoninos. We can no longer think about the added fermionic degrees of freedom as a device to construct a topological index. So, if the fermions are absent the supersymmetry is lost and with it the topological protection. This statement, however, is about the microscopic physics. It may very well be that the full non-linear phonon problem has the same low energy long distance physics as the corresponding supersymmetric non-linear phonon. After all, it is the physics close to the ground state which we used to build the model of balls and springs from constraint functions. It is therefore possible that the SUSY breaking terms are irrelevant in the renormalization group sense. If so, the topological protection may nevertheless emerge in the low energy limit. 

\subsection{What happens if the supersymmetry is broken?}
The models discussed in this paper are all ``toy'' models in that they correspond to useful simplifications of the actual phonon microscopic physics but in any real system, there will be perturbations that are not of the simple form. For example, the general quadratic phonon potential
\begin{equation}
  V_{phonon} = \frac{1}{2}\sum_{ij} u^{i\alpha}V_{i\alpha,j\beta}u^{j\beta} \neq \frac{1}{2} u^{i\alpha}A^T_{i\alpha,m}k^{m,n}A_{m,j\beta}u^{j\beta}
\end{equation} 
is not that of balls and springs ($V_{i\alpha,j\beta}\neq A^T_{i\alpha,m}k^{m,n}A_{m,j\beta}$). The supersymmetry can therefore be thought of as a symmetry emerging from the `toy-ness' of the model. We can speculate on what this might mean. The extent to which the toy model captures the full physics of a real material is likely the extent to which the supersymmetry will be obeyed. In particular, if supersymmetry demands gappless phonon edge modes in a system with gapped bulk phonon modes, these edge modes would likely be gapped by perturbations violating the toy-like property of balls and springs and the size of the gap is likely related to the strength of these non-toy-like perturbations.

\section{Topological aspects of magnons}
\subsection{Unfrustrated magnons}\label{sec:Magnons}
We can extend the above phonon discussion to magnons (or other similar systems) by recognizing its essential ingredient: constraints. The phonon problem is built on the local-in-configuration space condition that the extensions $e_m$ vanish in the ground state. Since a ground state also requires a vanishing momenta $p_{i\alpha}$, we see that the phonon problem is a quadratic form in the set of local-in-phase space constraints on the ground state. To extend to magnons, we therefore need models built around local-in-phase space constraints that define a ground state magnetic ordering pattern.

Perhaps the simplest example of such a constraint in a magnetic system is the vanishing total (classical) spin of a nearest neighbor bond in the N\'eel state of the square lattice antiferromagnet. Each nearest neighbor bond $m$ has a pair of sites $ij$ that satisfies $S_{m\alpha}\equiv S_{i\alpha} + S_{j\alpha} = 0$. Here each $S_{i\alpha}$ is a dynamical spin vector satisfying angular momentum Poisson brackets $\{S_{i\alpha},S_{i\beta}\} = \epsilon_{\alpha\beta}^{\gamma}S_{i\gamma}$ and lives in a spherical $2$-dimensional phase space. Such a condition is directly analogous to a motionless unstreatched spring with vanishing extension $e_m=0$ and momenta $p_{i\alpha}=0$, $p_{j\alpha}=0$. So we can follow our phonon discussion and use the condition $S_{m\alpha}=0$ to write down the SUSY charge
\begin{equation}
Q = \gamma^{m\alpha}S_{m\alpha}
\end{equation}
and SUSY Hamiltonian
\begin{multline}
H_{SUSY} = \frac{1}{2}\{Q,Q\} = \frac{1}{2} S_{m\alpha}\{\gamma^{m\alpha},\gamma^{n\beta}\}S_{n\beta} +\\ \frac{1}{2} \gamma^{m\alpha}\{S_{m\alpha},S_{n\beta}\}\gamma^{n\beta}
\end{multline}
Choosing $\{\gamma^{m\alpha},\gamma^{n\beta}\} = J^{m\alpha,n\beta}$ a positive definite symmetric matrix makes the first term a typical quadratic spin model with $S_{m\alpha}=0$ for a ground state (e.g. the case $J^{m\alpha,n\beta} = J\delta^{mn}\delta^{\alpha\beta}$ is just the Heisenberg model). This makes the $\gamma^{m\alpha}$ Grassmann variables obeying a Clifford algebra just like we found for the phonon problem. Additionally, $C_{m\alpha,n\beta}\equiv\{S_{m\alpha},S_{n\beta}\}$ follows from $\{S_{i\alpha},S_{i\beta}\}$ above and defines the fermion Hamiltonian again as Dirac's constraint matrix. We can then expand about a ground state to linear order using $S_{m\alpha} = A_{m\alpha,i\mu}x^{i\mu}$ with $x^{i\mu}\to(q^i,p_i)$ the phase space coordinates representing deviations of the spin on site $i$ from its ground state value. This renders $C_{m\alpha,n\beta}$ constant and decouples the two terms in $H_{SUSY}$. Linearizing allows us to demote the magninos described by $\gamma^{m\alpha}$ to fictitious particles whose use is to reveal the implications of the supersymmetry just like the case of the phoninos. However, the equations of motion of these magninos does not give us the ``square-rooted'' Hamiltonain for it is
\begin{equation}
  \dot\gamma^{m\alpha} = J^{m\alpha,m'\beta}C_{m'\beta,m''\gamma}\gamma^{m''\gamma}
\end{equation}
which, when squared, is not the magnon equations of motion. So the structure of the square-rooted Hamiltonian and its associated quantum mechanics-like supersymmetry doesn't apply here. Thus, by starting from the ground state condition $S_{m\alpha}=0$ we obtain the same structure as the phonon problem which presumably has a Witten index $\nu$ that can tell us about its zero modes without solving the equations of motion.

The above N\'eel state bond constraint $S_{m\alpha}$ is an application of the SUSY phonon problem to unfrustrated antiferromagnets. It readily generalizes to the ferromagnetic case which has a vanishing difference of spin vectors on each bond ($S_{m\alpha}\equiv S_{i\alpha} - S_{j\alpha} = 0$) provided we draw arrows on each bond to specify which of the two sites of nearest neighbor bond $m$ enters $S_{m\alpha}$ with a negative sign. It also readily generalizes to other bi-partite lattices such as the cubic lattice and the honeycombe lattice. Writing down the topological index expected for these generalizations will therefore allow us to determine if an analog of isostatic phonons exists and also how we may use it to study the loss of magnetic order. 

\begin{figure}
\includegraphics[width=\columnwidth]{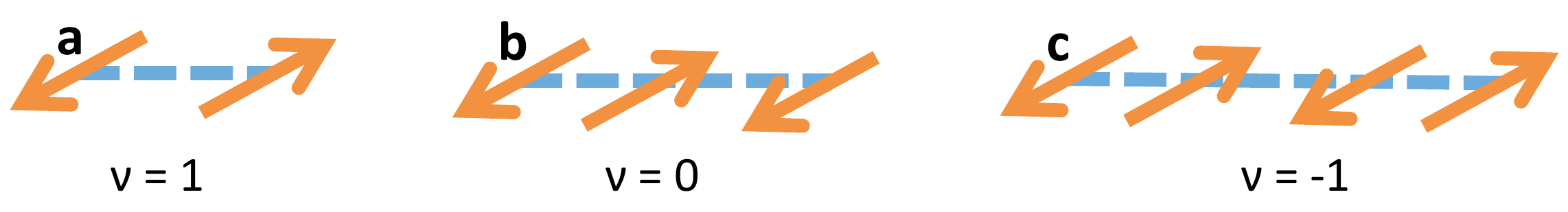}
\caption{Topological index for small chains with a colinear N\'eel ground state. Since $\nu$ becomes increasingly negative with increasing chain length going negative for four or more spins, essentially all unfrustrated magnetic systems will have a negative topological index. This explains the general robustness of magnetic order in many magnetic materials.}
\label{fig:unfrustratednu}
\end{figure}

For each case, we can guess the form of the topological index
\begin{align}\nonumber
 \nu_{unfrustrated} &= (\text{\# of magnons}) - (\text{\# of magninos}) \\
	& = 2N_s - 3N_b \\
	& = \text{nullity} \mat{A} - \text{nullity} \mat{A}^T
\end{align}
The last line here is not a guess, it follows from the previous line by the fundamental theorem of linear algebra as before. Presumably, if $\nu_{unfrustrated} > 0$, there must be a finite number of magnon zero modes and if it is negative a finite number of magnino modes. We will discuss the meaning and correctness of this statement below. But assuming it is correct for the present discussion, we see that essentially all unfrustrated magnets have $\nu_{unfrustrated} < 0$. Only dimerized systems which have two sites for each bond or trimerized systems with \emph{two bonds} for each trimer have non-negative $\nu_{unfrustrated}$ (see Fig. \ref{fig:unfrustratednu}). The best case with connected sites is the one dimensional chain with $N_s = N_b$ and $\nu_{unfrustrated} = -N_s$ for periodic boundary conditions. A similar best case holds for bond percolation on the square lattice where at threshold $N_b = N_s$\cite{Kesten1980}. So for unfrustrated magnets, the topological structure imposed by supersymmetry doesn't demand any magnon zero modes. 

The above derivation of the magnetic Maxwell-Calladine index was mostly a guess. To see that it is correct, we need to work out the supersymmetry relations between eigenmodes and the Witten index as we did for the phonon problem. This is a more complicated problem than the phonon case due to the existence of two types of zero modes\cite{Lawler2013} (canonical and gauge). But this just requires a little more organization and the following proceeds just like the phonon case.  

The equations of motion for both bosonic and fermionic problems are
\begin{eqnarray}
\dot x^{i\mu} &= \sigma^{i\mu,j\nu} A^T_{j\nu,m\alpha}J^{m\alpha,m'\beta}A_{m'\beta,k\lambda}x^{k\lambda}\\
\dot \gamma^{m\alpha} & = J^{m\alpha,m'\beta}A_{m'\beta,i\mu}\sigma^{i\mu,j\nu}A^T_{j\nu,m''\gamma}\gamma^{m''\gamma}
\end{eqnarray}
where $\sigma^{i\mu,j\nu}=\{x^{i\mu},x^{j\nu}\}= \delta^{ij}\epsilon^{\mu\nu}$ is the Poisson bracket tensor with $\epsilon^{\mu\nu}$ the two dimensional Levi-Civita tensor and we expanded the constraint matrix to leading order (where $C_{m\alpha,m'\beta} = A_{m\alpha,i\mu}\sigma^{i\mu,j\nu}A^T_{j\nu,m'\beta}$). Lets identify first the types of bosonic eigenmodes, then the types of fermionic eigenmodes and then see how they map into each other by sypersymmetry.

For the $x^{i\mu}$ equations of motion, we can break up the set of eigenmode observables $v_{i\mu}\dot x^{i\mu}$ into four types of modes.
\begin{enumerate}
\item Finite frequency modes obeying $v_{i\mu}\sigma^{i\mu,j\nu} A^T_{j\nu,m\alpha}J^{m\alpha,m'\beta}A_{m'\beta,k\lambda} = i\omega v_{k\lambda}$ with $\omega>0$;
\item Gauge zero modes that are eigenmodes with $\omega=0$ but whose observable $v_{i\mu}x^{i\mu}$ commutes with all other observables corresponding to other $\omega=0$ eigenmodes. These obey $A_{m\alpha,i\mu}\sigma^{i\mu,j\nu}v_{j\nu}=0$ and $\{v_{i\mu}x^{i\mu},w_{j\nu}x^{j\nu}\}=0$ where $w_{j\nu}$ is any vector satisfying $A_{m\alpha,i\mu}\sigma^{i\mu,j\nu}w_{j\nu}=0$.
\item Canonical zero modes that are eigenmodes with $\omega=0$ that are not gauge zero modes. These obey $A_{m\alpha,i\mu}\sigma^{i\mu,j\nu}v_{j\nu}=0$ and $\{v_{i\mu}x^{i\mu},w_{j\nu}x^{j\nu}\}\neq 0$ for some $w_{j\nu}$ satisfying $A_{m\alpha,i\mu}\sigma^{i\mu,j\nu}w_{j\nu}=0$.
\item Non-eigenmodes of the equations of motion that can arise because the eigenvalue problem here is a non-symmetric matrix that does not guarantee a complete set of eigenvectors. These turn out to be the modes conjugate to the gauge zero modes. We will therefore call them "conjugate gauge modes".
\end{enumerate} 
These modes are best understood by choosing a canonical basis that pairs them up into position $Q_a=v_{i\mu}\dot x^{i\mu}$ and momentum $P_a=w_{j\nu}x^{j\nu}$ observables. Then each pair has Hamiltonian
\begin{equation}\label{eq:modeform}
  \frac{P_a^2}{2m_a} + \frac{1}{2}m_a\omega_a^2Q_a^2
\end{equation}
In this light, finite frequency modes have $m_a\neq 0$, $\omega_a\neq 0$, canonical zero modes have $m_a=\infty$, $\omega_a=0$ and $m_a\omega_a^2=0$ and gauge modes have $m_a>0$, $\omega_a=0$. 

For the $\gamma^{m\alpha}$ equations of motion, we can break up the set of eigenmodes $g_{m\alpha}\gamma^{m\alpha}$ into:
\begin{itemize}
\item finite frequency modes obeying $g_{m\alpha}J^{m\alpha,m'\beta}A_{m'\beta,i\mu}\sigma^{i\mu,j\nu}A^T_{j\nu,m''\gamma} = i\omega g^{m''\gamma}$;
\item self field modes obeying $A^T_{i\mu,m\alpha}J^{m\alpha,m'\beta}g_{m'\beta}=0$;
\item magnino gauge modes that do not satisfy $A_{m\alpha,m'\beta}J^{m'\beta,m''\gamma}g_{m''\gamma}=0$ but are eigenmodes with $\omega=0$.
\end{itemize}
Self field modes are modes that correspond to a choice of local magnetic fields on the bonds $h^{m\alpha}$ that contribute no Zeeman energy
\begin{equation}
 H_{Zeeman} =  - h^{m\alpha}S_{m\alpha} = 0
\end{equation}
They identify a redundancy in the set of functions $S_{m\alpha}$ when viewing them as constraints on the ground state configuration. The ``magnino gauge modes'' are not gauge modes in the sense of zero modes whose conjugate is not a zero mode. Their name here comes from their use in counting gauge modes: Dirac found\cite{Dirac1950,Dirac1958} there is a gauge mode for every eigenvector in the nullspace of the constraint matrix that does not correspond to a redundant constraint (i.e. is not self field mode). 

\begin{figure}
\includegraphics[width=\columnwidth]{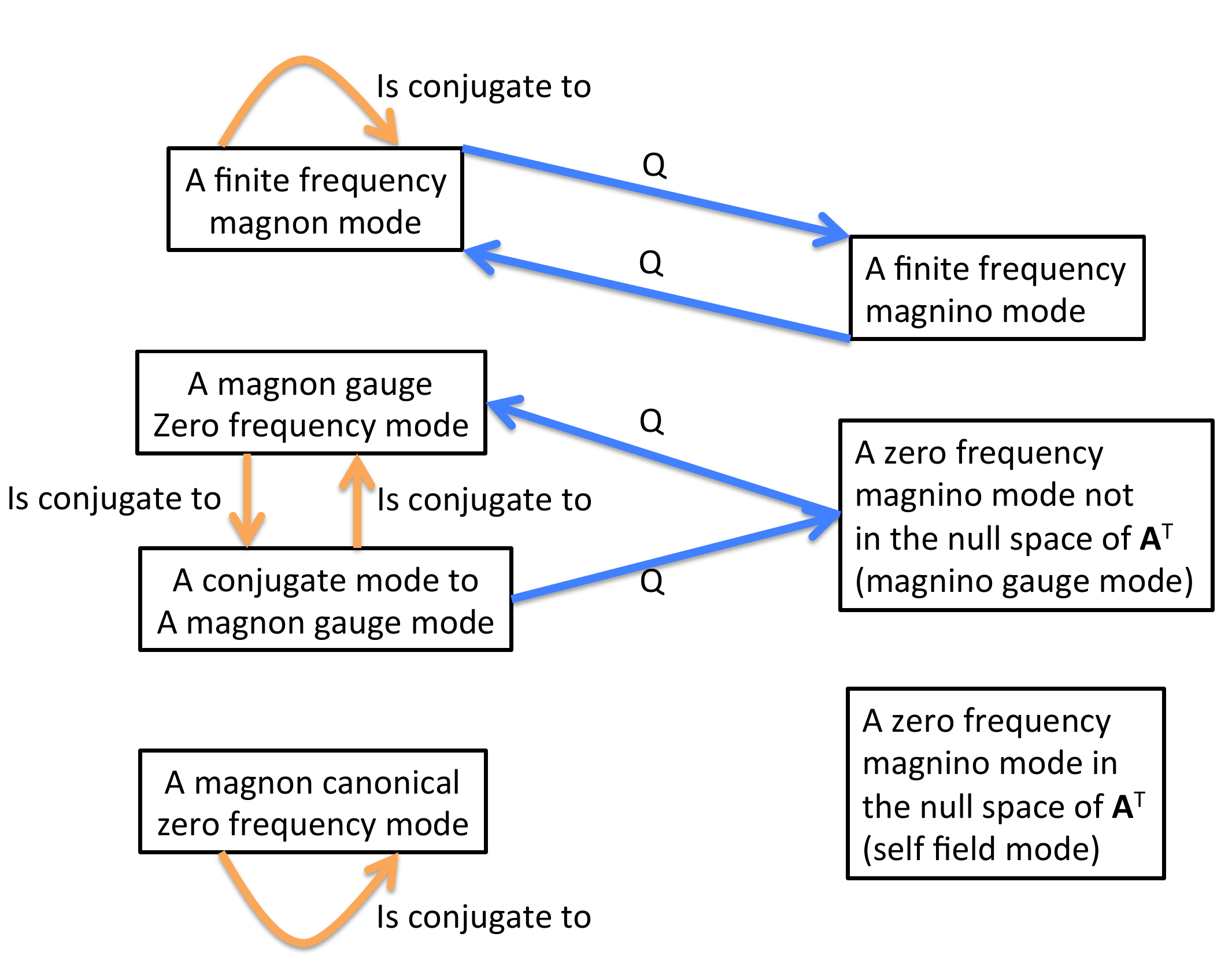}
\caption{A graphical depection of the relationships between eigenmodes of the magnon-magnino supersymmetric model. The magnon modes are broken down into four groups due to the canonical structure of the zero modes: some zero modes have a canonical conjugate (canonical zero frequency mode) and others do not (gauge zero frequency mode). Since each arrow maps a unique eigenmode to a unique eigenmode, the finite frequency magnon and magnino modes are isomorphic. Also the magnon gauge zero frequency modes are isomorphic to the magnino ``gauge'' zero frequency modes which explains Dirac's use of the constraint matrix to count gauge modes purely from supersymmetry. Note: like Fig. \ref{fig:PhononQMap}, this figure is also an olog\cite{Spivak2014}.}
\label{fig:FMagnonsQMap}
\end{figure}

Now with all these modes identified, we can pass them through the supersymmetry map $\{\cdot,Q\}$ and see how they are related by supersymmetry. The result is presented in Fig. \ref{fig:FMagnonsQMap}. Since no two eigenmodes mapped by $\{\cdot,Q\}$ maps to the same eigenmode, we must have:
\begin{itemize}
\item the number of finite frequency magnon modes $N_>$ is equal to the number of finite frequency magnino modes;
\item the number of magnon gauge modes $N_G$ is equal to the number of magnino ``gauge'' modes.
\end{itemize}
In addition, we see that the supersymmetry map gives us a direct explanation of Dirac mode counting: magnino gauge modes map to magnon gauge modes under supersymmetry and the conjugate to this magnon gauge mode maps back to the magnino gauge modes. This defines an isomorphism between magnon and magnino gauge modes and so the number of each must be the same. The supersymmetry map also verifies our guess for the single particle Witten index/magnetic Maxwell-Calladine index for the above conditions on the number of modes in each set implies the difference between the number of magnon and magnino modes is
\begin{align}
\nu &= \underbrace{2N_s}_\text{$x^{i\mu}$ modes} - 
         \underbrace{3N_b}_\text{$\gamma^m$ modes}\\
      &= (N_{>} + 2N_G + 2N_c) - (N_{>} + N_G + N_{sf})\\
      &=  N_G + 2N_c - N_{sf}
\end{align}
where $N_c$ is half the number of zero modes with a conjugate pair that is also a zero mode and $N_{sf}$ is the number of self field modes. And so our guess of Eq. \ref{eq:Aindex} is indeed correct with a positive $\nu$ demanding $N_G + 2N_c\geq \nu$ and a negative $\nu$ demanding $N_{sf} \geq |\nu|$. 

We can also determine the topological another way. We see from the above discussion that $N_G+2N_c = \text{nullity}\mat{A}$ (dimension of the constraint surface $S_{\Delta\alpha}=0$ in phase space) and $N_{sf} = \text{nullity}A^T$ since these null spaces define the modes that do not get mapped under supersymmetry. Hence, the Witten index in the magnon problem also obeys $\nu = \text{nullity}\mat{A} - \text{nullity}A^T$. 

So our guess for the topological index is correct but this in turn implies that unfrustrated magnons are always far away from the analog of mechanical collapse expected for $\nu>0$ (i.e. they all have a negative $\nu$). So, to explore the physics of the Kane-Lubensky theory in magnetic systems, we need to generalize beyond unfrustrated magnetic systems. But before doing so, we turn to an illustrative example of the above magnon and magnino eigenvalue problems: the simple ferromagnet on the square lattice. 

\subsection{SUSY in the square lattice ferromagnet}
We can write the Hamiltonian for the square lattice ferromagnet Heisenberg model as
\begin{equation}
  H = J\sum_{\langle ij\rangle}\vec S_i\cdot \vec S_j = \frac{1}{2} S_{m\alpha}J^{m\alpha,n\beta}S_{n\beta}
\end{equation}
where $J^{m\alpha,n\beta} = J\delta^{mn}\delta^{\alpha\beta}$, $m$ labels nearest neighbor bonds $\langle ij\rangle$ and $\alpha$ the spin vector components. The ground state therefore satisfies $S_{m\alpha} \equiv S_{i\alpha}-S_{(i+\hat x)\beta} = 0$ on each horizontal bond and $S_{m\alpha} \equiv S_{i\alpha}-S_{(i+\hat y)\beta} = 0$ on each vertical bond. Up to global spin rotations, this uniquely selects the ground state to be the uniform state. 

To linearize around a ground state, let's choose the case $S_{ix} = 1$, $S_{iy} = S_{iz} = 0$. Expressing the spin vectors using canonical coordinates $\vec S_i = (\cos(q^i)\sqrt{1-p_i^2},\sin(q^i)\sqrt{1-p_i^2},p_i)$ with $\{q^i,p_i\}=1$ so that $\{S_{i\alpha},S_{i\beta}\} = \epsilon_{\alpha\beta}^\gamma S_{i\gamma}$ we can linearize by expanding in $q^i$ and $p_i$. This gives $\vec S_{i} = (1, q^i, p_i)$. Similarly, we can expand the constraint $S_{m\alpha} = A_{m,i\mu}x^{i\mu}$ and identify $x^{iq} = q^i$ and $x^{ip} = p_i$ with $\mu = q,p$ indexing to which canonical variable $x^{i\mu}$ refers. So we find
\begin{equation}
  A_{mx,i\mu} = 0, \quad A_{my,i\mu} = \pm\delta_{\mu,q},\quad
 A_{mz,i\mu} = \pm\delta_{\mu,p}
\end{equation}
where the sign depends on which spin $i\mu$ refers to in the expression $S_{m\alpha} = S_{i\alpha}-S_{j\beta}$. 

Now since the ground state is periodic with the lattice, we may Fourier transform. To do this, let $m = \vec R + \vec d$ with $\vec d = a\hat x/2$ or $\vec d = a\hat y/2$ denoting the center of each bond and $i = \vec R$ denoting the site of the lattice. Then $A_{m,i\mu} = A_{\vec R+\vec d,\vec R'\mu}$ and we see its translational symmetry by just shifting both  $\vec R$ and $\vec R'$ by $\vec a$ and for some lattice vector $\vec a$. We can then expand $A_{m,i\mu}$ in Fourier components by
\begin{equation}
  A_{\vec R +\vec d\alpha,\vec R'\mu} = 
\sum_{\vec k} A_{\vec d\alpha,\mu}e^{i\vec k\cdot(\vec R-\vec R')}/N_u
\end{equation}
where $N_u$ is the number of unit cells. This introduces the 6x2 matrix
\begin{multline}
  {\bf A}^T(\vec k) =\\ 
\begin{pmatrix}
 0 & 1-e^{ik_xa} & 0 & 0 & 1-e^{ik_ya} & 0\\
0 & 0 & 1-e^{ik_xa} & 0 & 0 & 1-e^{ik_ya} 
\end{pmatrix}
\end{multline}
and allows us to express the magnon Hamiltonian as
\begin{align}
  H &= \frac{1}{2} \sum_\vec k {\bf x}^T(-\vec k){\bf A}^T(-\vec k){\bf \tilde J}(\vec k){\bf A}(\vec k){\bf x}(\vec k)\\
  & \equiv \frac{1}{2} {\bf x}^T(-\vec k) \bf h(\vec k) {\bf x}(\vec k)
\end{align}
Multiplying the matrices out for the nearest neighbor Heisenberg model (${\bf J}$ a 6x6 identity matrix) we find ${\bf h}(\vec k) = 4(\sin^2(k_xa) + \sin^2(k_ya)){\bf I}_{2x2}\equiv \omega(\vec k){\bf I}_{2x2}$ where ${\bf I}_{2x2}$ is the 2x2 identity matrix giving a quadratic dispersion for small $\vec k$ as expected.

Now we can identify mode types as finite frequency, magnon gauge, conjugate to magnon gauge and magnon canonical zero mode. At $\vec k=0$ we find a zero mode with ${\bf h}(\vec 0) = 0$. It is therefore of the form of Eq. \ref{eq:modeform} with $m_a\to\infty$, $\omega_a\to0$ and $m_a\omega_a^2\to0$. Hence it is a canonical zero mode. For finite $\vec k$, we see that Hamiltonian is of the form of Eq. \ref{eq:modeform} with $m_a=1/\omega(\vec k)$ and $\omega_a=\omega(\vec k)$ finite. Hence these are finite frequency magnon modes. There are unfortunately no magnon gauge modes are conjugate gauge modes to illustrate this point.

The magnino Hamiltonian takes a similar form. It is
\begin{align}
  H_{magnino} &= \frac{1}{2} \gamma^T(-\vec k){\bf A}(-\vec k){\bf \tilde \sigma}(\vec k){\bf A}^T(\vec k)\gamma(\vec k)\\
  &\equiv \frac{1}{2} \gamma^T(-\vec k){\bf C}(\vec k)\gamma(\vec k)
\end{align}
where $\sigma(\vec k) =i\sigma_y$ is the antisymmetric Poisson bracket tensor. Now ${\bf C}(\vec k)$ is a $6x6$ antihermetian matrix:
\begin{equation}
 {\bf C}(\vec k) = \begin{pmatrix}
  0 & 0 & 0 & 0 & 0 & 0\\
  0 & 0 & \xi_x^*\xi_x & 0 & 0 & \xi_x^*\xi_y\\
  0& -\xi_x^*\xi_x & 0 & 0 & -\xi_x^*\xi_y & 0\\
  0 & 0 & 0 & 0 & 0 & 0\\
  0 & 0 & \xi_x\xi_y^* & 0 & 0 & \xi_y^*\xi_y\\
  0 & -\xi_x\xi_y^* & 0 & 0 & -\xi_y^*\xi_y & 0
\end{pmatrix}
\end{equation}
Choosing ${\bf J}$ a 6x6 identity matrix so that $\gamma(\vec k)$ obey canonical commutation relations then allows us to diagonalize this matrix. The result is four zero eigenvalues, $i\omega(k)$ and $-i\omega(k)$. Hence the finite frequency modes have exactly the same eigenvalues. 

Again we can distinguish the different mode types though here we haven't expressed a simple canonical form for the diagonalized fermion Hamiltonian. We don't have Eq. \ref{eq:modeform} to help guide us. But we can work directly with the definitions of the modes. Clearly finite frequency modes magnino modes correspond to the eigenvalues $i\omega(k)$ and $-i\omega(k)$ for $\vec k\neq 0$ and there are exactly the same number of these as finite frequency magnon modes. The four zero eigenvalues and the case $\omega(\vec k=0)=0$ fall into the null space of ${\bf A^T}$ and are "self field modes". There are no zero frequency modes not in the null space of ${\bf A}^T$ so there are no "magnino gauge modes" consistent with the absence of magnon gauge modes discuss above. 

\subsection{Frustrated magnons}\label{sec:FrustratedMagnons}
Perhaps one way to overcome the ubiquitous negative $\nu$ of unfrustrated magnets is to generalize to other degrees of freedom. For this to improve upon the unfrustrated case, we need to increase the number of degrees of freedom on each site compared to the number of constraints on each bond. This seems difficult for magnetic systems for each site naturally has its own constraints (such as unit length spin vectors) in addition to the bond constraints. For example, consider $O(2)$ quantum rotors each characterized by $O(2)$ spin unit vector $s_{i\alpha}$, ($s_{ix} = \cos\theta_i$, $s_{iy}=\sin\theta_i$) and angular momenta $L_i$ conjugate to $\theta_i$ (see Sachdev's book\cite{Sachdev2011}). In this case $L_i=0$ and $s_{m\alpha} = s_{i\alpha}\pm s_{i\beta}$ in the ground state so we can define $Q = \Gamma^i L_{i} + \gamma^{m\alpha}s_{m\alpha}$. Such a $Q$ leads to $H_{SUSY} = H_{quantum rotor} + H_{fermion}$ with fermions again decoupling in the linearized limit. But now our guess for the topological index is
\begin{equation}
\nu_{rotor} = 2N_s - N_s - 2N_b=N_s - 2N_b
\end{equation}
which is typically even more negative than $\nu_{unfrustrated}$ (and the situation doesn't improve for $O(3)$ rotors). But notice that here we kept the same number of phase space degrees of freedom on each site and just transferred one of the bond constraints to a site constraint. Thus it is even more difficult to introduce magnon zero modes for quantum rotors than for quantum spins and they do not seem to be a viable solution.

Let us then turn to changing the constraint functions entirely. Lets replace the vanishing bond constraint $S_{m\alpha}=0$ with a vanishing total spin on a plaquette constraint $S_{p\alpha}=0$. Such a constraint arises in the nearest neighbor triangular, kagome, pyrochlore antiferromagnets and the $J_1$-$J_2$ square lattice antiferromagnet with $J_2 = J_1/2$. All of these cases fall in to the category of frustrated magnets. For example, the triangular lattice Heisenberg model can be written
\begin{equation}
  H = J\sum_{\langle ij\rangle}S_{i\alpha}\delta^{\alpha\beta}S_{j\alpha} = \frac{J}{4}\sum_{\Delta\alpha} S_{\Delta\alpha}^2 + const
\end{equation}
where $\Delta$ labels each triangle of the triangular lattice and the $J_1$-$J_2$ square lattice antiferromagnet at $J_2=J_1/2=J/2$ can be written 
\begin{equation}
  H = J\sum_{\langle ij\rangle}S_{i\alpha}\delta^{\alpha\beta}S_{j\alpha} = \frac{J}{4}\sum_{\Box\alpha} S_{\Box\alpha}^2 + const
\end{equation}
We can again guess the topological index. Here it is
\begin{equation}
\nu_{frustrated} = 2N_s - 3N_p
\end{equation}
where $3N_p$ is the number of plaquette constraints (and fermion Grassmann numbers). This gives for $N_u$ unit cells and periodic boundary conditions $\nu_{frustrated} = -4N_u$ for the triangular lattice (2 triangles, 1 site in unit cell), $\nu_{frustrated} = -N_u$ for the $J_2=J_1/2$ square lattice case (1 square, 1 site in unit cell), $\nu_{frustrated} = 0$ for the kagome case (2 triangles, 3 sites in unit cell) and $\nu_{frustrated} = 2N_u$ for the pyrochlore lattice (2 tetrahedrons, 4 sites in unit cell). So frustration enables a wide variety of models to explore the impact of the full range of $\nu$ values on the ordering tendencies of the spins. We next work out the special $\nu=0$ kagome case as a concrete example and use it to clarify a number of the above general statements. 

On the kagome lattice (and indeed many highly geometrically frustrated magnets\cite{Moessner1998a,Moessner1998b}) the Hamiltonian can be written
\begin{equation}\label{eq:HFrustratedMagnons}
  H_{kagome} = \frac{1}{2}S_{\Delta\alpha}J^{\Delta\alpha,\Delta'\beta}S_{\Delta'\beta}
\end{equation}
where $S_{\Delta\alpha} = S_{i\alpha}+S_{j\alpha}+S_{k\alpha}$ is the total spin on triangle $ijk$. For the case, $J^{\Delta\alpha,\Delta'\beta}= J\delta^{\Delta,\Delta'}\delta{\alpha\beta}$, this model reduces to the nearest neighbor Heisenberg model on a lattice of corner sharing triangles such as the kagome lattice and hyperkagome lattice. We will therefore assume $J^{\Delta\alpha,\Delta'\beta}$ has positive definite eigenvalues and defines a metric. 

This model has the feature that the ground states all satisfy the local condition $S_{\Delta\alpha}=0$. Using this constraint, we can again write down a supersymmetric charge
\begin{equation}
  Q = \gamma^{\Delta\alpha}S_{\Delta\alpha}
\end{equation}
and compute the corresponding supersymmetric Hamiltonian
\begin{equation}
  H_{SUSY} = \frac{1}{2} S_{\Delta\alpha}\{\gamma^{\Delta\alpha},\gamma^{\Delta'\beta}\}S_{\Delta'\beta} + 
\frac{1}{2} \gamma^{\Delta\alpha}C_{\Delta\alpha,\Delta'\beta}\gamma^{\Delta'\beta}
\end{equation}
So we need to choose $\{\gamma^{\Delta\alpha},\gamma^{\Delta'\beta}\}=J^{\Delta\alpha,\Delta'\beta}$ to make the first term $H_{kagome}$. The second term is the magnino Hamiltonian and like the phonino Hamiltonian is determined by the constraint matrix ($C_{\Delta\alpha,\Delta'\beta} = \{S_{\Delta\alpha},S_{\Delta'\beta}\}$). Hence, we obtain the same structure as we did for the unfrustrated case. We just replace the index $m$ with $\Delta$.

We can now compute the topological index for kagome antiferromagnets. After linearizing as before the analog of the magnetic Maxwell-Calladine index is now
\begin{equation}\label{eq:Aindex}
  \nu \equiv 2N_s - 3N_\Delta  = N_G + 2N_c - N_{sf}
\end{equation}
This is different from the unfrustrated case only in replacing $N_b$ the number of bonds with $N_\Delta$ the number of triangles. For periodic boundary conditions, we see that $\nu/N_u = 2n_s - 3n_\Delta=0$ where $N_u$ is the number of unit cells considered, $n_s = 3$ is the number of sites within a unit cell and $n_\Delta=2$ is the number of triangles within a unit cell. We additionally see this holds for other kagome lattices like the hyperkagome lattice of Na$_4$Ir$_3$O$_8$\cite{Okamoto2007} in three dimensions which has $n_s = 12$, $n_\Delta=8$. So the kagome family of lattices all have $\nu=0$. 

\begin{figure}
\includegraphics[width=0.45\columnwidth]{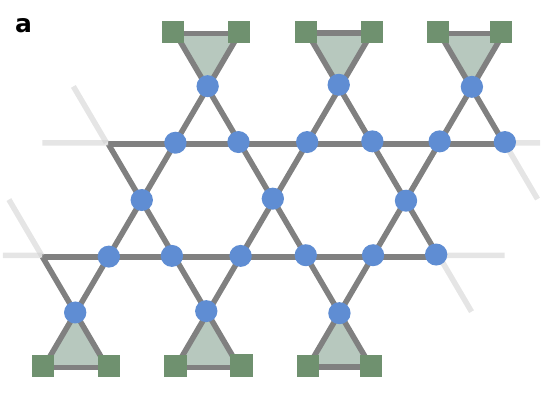}
\includegraphics[width=0.45\columnwidth]{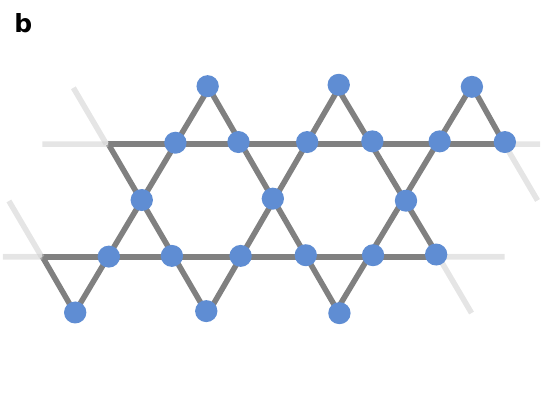}
\caption{Two examples of open boundary conditions for the kagome lattice. Each of these cases has $\nu=2N_s-3N_\Delta>0$. a) has dangling triangles and $\nu=2(33)-3(18)=12$, b) has no dangling triangles and $\nu=2(21)-3(12)=6$. It appears that $\nu>0$ for any open boundary conditions and only reaches zero for periodic boundary conditions. Thus we can always gap out all bulk modes with a suitable choice of the matrix $\mat{A}$ but not so at the edge. A magnon gauge or canonical zero mode must exist on the edge so long as the magnon Hamiltonian is of the form Eq. \ref{eq:HFrustratedMagnons}.}
\label{fig:KagomeEdgeStates}
\end{figure}

Given that $\nu=0$ for periodic boundary conditions, lets see what the index demands for various open boundary conditions.  For the two finite kagome clusters shown in Fig. \ref{fig:KagomeEdgeStates}, we have $\nu = 12$ and $\nu=6$. So either type of boundary condition will have $\nu> 0$ and demand zero modes. Presumably, this holds for any boundary condition that still allows us to write the Hamiltonian in the general form of Eq. \ref{eq:HFrustratedMagnons}. However, other boundary conditions may include a bond constraint function $S_{m\alpha}$ in place of a triangle constraint function $S_{\Delta\alpha}$ and we would need to rethink the problem and be sure these constraints really do define ground state configurations (i.e. it could be that there are no spin configurations that satisfy all the constraints). In these cases, it may be that $\nu < 0$ for such boundaries given that we found bond constraints negatively contribute to $\nu$ so that $\nu$ doesn't demand any zero modes. Hence, it may depend on the nature of the boundary conditions as to whether zero modes are expected. However, there exist boundary conditions where the topological index will demand the existence of zero modes even in the absence of any spin rotational symmetry breaking (as it does in systems with $\nu>0$ for periodic boundary conditions such as the nearest neighbor pyrochlore antiferromagnet). 

The above results show that indeed we can extend the supersymmetry of the linearized phonon problem to both unfrustrated and frustrated magnons. In addition, we find that the kagome family is an analog of ``Maxwell lattices''\cite{Lubensky2015}. It has $\nu=0$ for periodic boundary conditions and $\nu>0$ at least for certain open boundary conditions. If we gap out all modes, with supersymmetry allowed perturbations, we would then have a magnetic analog of an isostatic lattice (a Maxwell lattice with no zero modes). 

\section{Phonon physics in magnon systems}
\begin{figure}
\includegraphics[width=\columnwidth]{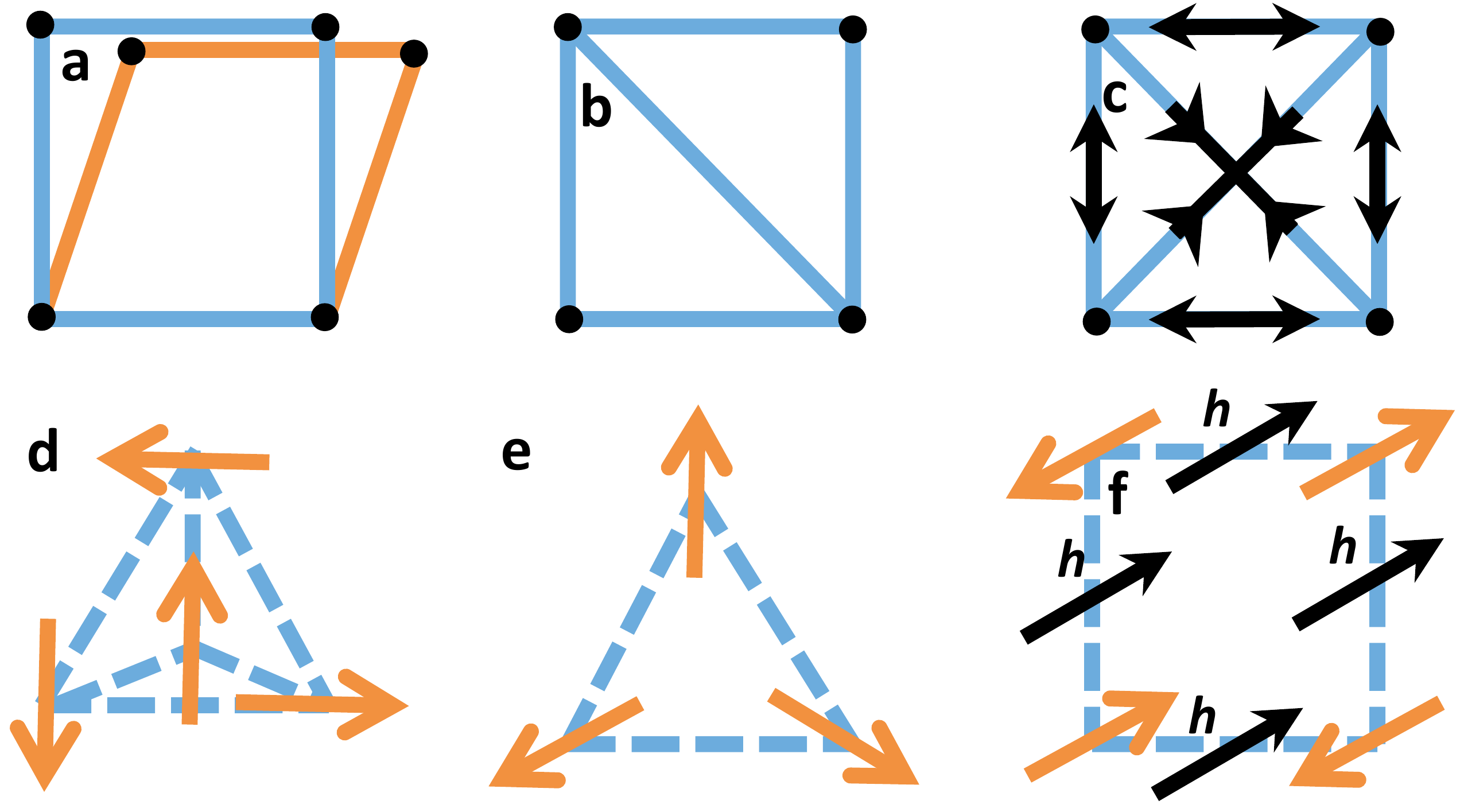}
\caption{Comparison between mechanical stability of beams connected by pins and magnetic stability in small systems. Beams are colored blue, pins colored black, dashed blue lines represent neighboring sites and orange arrows represent spins. {\bf a} a square set of beams with one zero mode shown shown using orange beams. {\bf b} a square set of beams with one cross beam that eliminates the zero mode. {\bf c} a square set of beams with two cross beams that enables a state of self stress (with tensions shown via black arrows). {\bf d} Magnetic analog of {\bf a} when imposing a vanishing total spin constraint on the tetrahedron: here each pair of spins pointing in opposite directions can be rotated independent of the other pair of spins. {\bf e} Magnetic analog of {b} when imposing vanishing total spin on a triangle. This has no redundant constraints or zero modes aside from global spin rotations. {\bf f} Magnetic analog of {\bf c} when imposting vanishing total spin on each bond expected in unfrustrated antiferromagnets. It supports several self field modes including the uniform magnetic field ${\bf h}$ applied to each neighboring spins as shown with black arrows (equivalent to a uniform field on each spins).}
\label{fig:isomagnetism}
\end{figure}

Let us then turn to illuminating some of the new physics that is transported from the theory of phonons into the theory of magnetism by our formal developments. Perhaps most striking is the existence of a new phase of magnetism which is on the verge of magnetic collapse that we will here refer to as isostatic magnetism. We can understand the existence of this phase in simple terms as shown in Fig. \ref{fig:isomagnetism}. First consider the three different mechanical structures, the square, the square with one diagonal beam and the square with two diagonal beams we see three different forms of rigidity: floppy, fragile, and firm. The square is floppy because it has one zero mode. The square with one diagonal brace is fragile because though it is rigid, if you cut/break any of the braces, it becomes floppy. The square with two diagonal braces is firm because it remains rigid if you cut/break any one of the braces. 

Now consider the three magnetic structures, four spins on a square, three spins on a triangle and four spins on a tetrahedron in the figure. In the square case, we subject the spins to $S_{ma} \equiv S_{ia} + S_{ja} = 0$ for $i$, $j$ nearest neighbor, $m\leftrightarrow ij$ indexes bonds and $a = x,y,z$. In the triangle and tetrahedron, we merely impose that the total spin of the object is zero. The square is the analog of the firm mechanical structure: after imposing $S_{ia} + S_{ja} = 0$ on three of its four sides, the fourth side is redundant. Removing one bond would not destroy the N\'eel order of its ground state. The triangle is the example of the fragile mechanical structure. All three constraints $S_{1a} + S_{2a} + S_{3a}=0$ that define its ground state are needed to set the spins to be 120$^o$ apart (up to global spin rotations). If we drop one of these three constraints, it gains a zero mode. Finally, the tetrahedron is the example of the floppy mechanical structure. We can choose $S_{1a} = -S_{2a}$ and $S_{3a}=-S_{4a}$ independently and still $S_{1a}+S_{2a}+S_{3a}+S_{4a}=0$. In this light, the new isostatic magnetic order that we can transport from the theory of phonons, is an order like the triangle case: if we were to drop any one term in the Hamiltonian that imposes a constraint on the ground state, there is no redundancy and the system becomes floppy. Isostatic magnetism is fragile.

Remarkably, since we have already the other two cases, ``floppy'' magnetism in the form of highly frustrated magnets and ``firm'' magnetism for most other magnetic systems, it may just be a matter of time to identify isostatic magnetism at least within some temperature range. One group of materials that may contain this novel magnetism is the family of distorted kagome antiferromagnets. An  undistorted ideal kagome antiferromagnet has $\nu=0$ but many zero modes. Adding distortions could then lift the ground state degeneracy without changing $\nu$ since distortions do not change the topology of the lattice network. Since distortioned kagome lattices are much more common than ideal kagome lattices, there may already be a number of such systems that are isostatic magnets.

Another physical phenomena we can transport from the phonon theory is the existence of self field modes that are the analog of self stress modes. The square with two diagonal beams of Fig. \ref{fig:isomagnetism} has such a self stress mode because of its redundancy of constraints. If one puts the diagonal beams in tension $T$, then one can place the horizontal and vertical beams in tension $-T/\sqrt{2}$ and no force will be exerted on the pins at is vertices. Similarly, one can add a ``bond magnetic field'' $h^{ma}$ via a term in the Hamiltonian $\delta H = - h^{ma}S_{ma}$ and there must exist $h^{ma}$ such that no net field is experienced by a spin to linear order in the displacements of the spins from their ground states. We know actually that there are at least $N_{sf} = 6$ of these from the topological index which here is $\nu = 2N_c+N_G - N_{sf} = 2N_s - 3N_b = 2\times 4 - 3\times 4=-4$. Because we include spin rotation invariance, there are two zero modes and so $N_0 - N_{sf} = -4$ demands $N_{sf}=-6$. If $m$ labels bonds cyclically around the square, three of these six modes span $h_{1a} = -h_{2a} = h_{3a} = - h_{4a}$. These produce $\delta H = 0$ non-linearly (i.e. independent of setting $S_{ma} = A_{ma,i\mu}x^{i\mu}$). Of the remaining three, one is the case $h_{1z} = h_{2z} = h_{3z} = h_{4z}\equiv h_z$ assuming we choose a Neel ground state in the $z$-direction. This case produces
\begin{equation}
  \delta H = h_z \sum_m S_{mz} = 2 h_z \sum_ i S_{iz} = 0
\end{equation}
only after we linearize the spins $S_{ia}$ around their magnetic ground state by setting $S_{ma} = A_{ma,i\mu}x^{i\mu}$. To non-linear order in the deviations of the spins from their ground state, this case will produce a non-zero energy $\delta H \neq 0$. The remaining two self field modes behave similarly. So self field modes exist and extend our understanding of the theory of magnetism.

Actually, we can understand what an experimentalist could achieve with a knowledge of self field modes in analogy with the self stress modes. Self stress modes have been studied at least as far back as Eshelby's 1957 paper\cite{Eshelby1957} who emphasized the engineering importance of self stresses in an elastic medium with inhomogeneities. More recently they have been understood as a means of selective mechanical failure of metamaterials\cite{Paulose2015}. In a $\nu=0$ lattice with one self stress mode and one zero mode located in different regions, a load was shown to cause mechanical failure exactly at the self stress mode bonds. Similarly, an applied field that corresponds to a self field mode should cause a spin flip. Indeed, in the case above for the square with a field $h_{mz} = h_z$ applied, we see that the spins will flip due to the non-zero energy $\delta H\neq 0$ caused by the non-linearities. The other two self field modes not discussed in detail similarly cause a spin flip, just not all spins flip in those cases. So such self field modes correspond to spin flip transitions and their identification in the constraint matrix eigenvalue problem may enable the discovery of a variety of such spin flip transitions. 

\section{Discussion} 
In this paper, we have recognized a supersymmetry in the structure of Hamiltonians that energetically impose local constraints on their ground states. This allowed us to formally connect recent development in the theory of phonons to a wider range of systems including frustrated and unfrustrated magnets.  The connection ultimately arose from the similar role of the vanishing extension of springs $e_m=0$ of phonon ground states to the vanishing total spin on a bond in unfrustrated magnets and vanishing total spin on a triangle, square, tetrahedron or other local simplex in frustrated magnetic systems. It also highlights the role of Dirac's constraint matrix as a central pillar of these recent developments as well as proves that a topological index exists in all these systems.  Lets end our discussion of these results with some final comments.

Remarkably, this supersymmetry identifies Maxwell's counting of degrees of freedom $D = N - K$, where $N$ is the total number of degrees of freedom and $K$ is the total number of local constraints imposed by the Hamiltonian (i.e. possibly a redundant set), to the Witten index $\nu$ associated with the supersymmetry. This Maxwell counting, though long used for balls and springs and other mechanical structures, was proposed by Moessner and Chalker as a means of characterizing highly frustrated antiferromagnets\cite{Moessner1998a,Moessner1998b}. Here we have connected their work with the actual number of zero modes through recognizing it as the topological index. 

\begin{figure}
\includegraphics[width=\columnwidth]{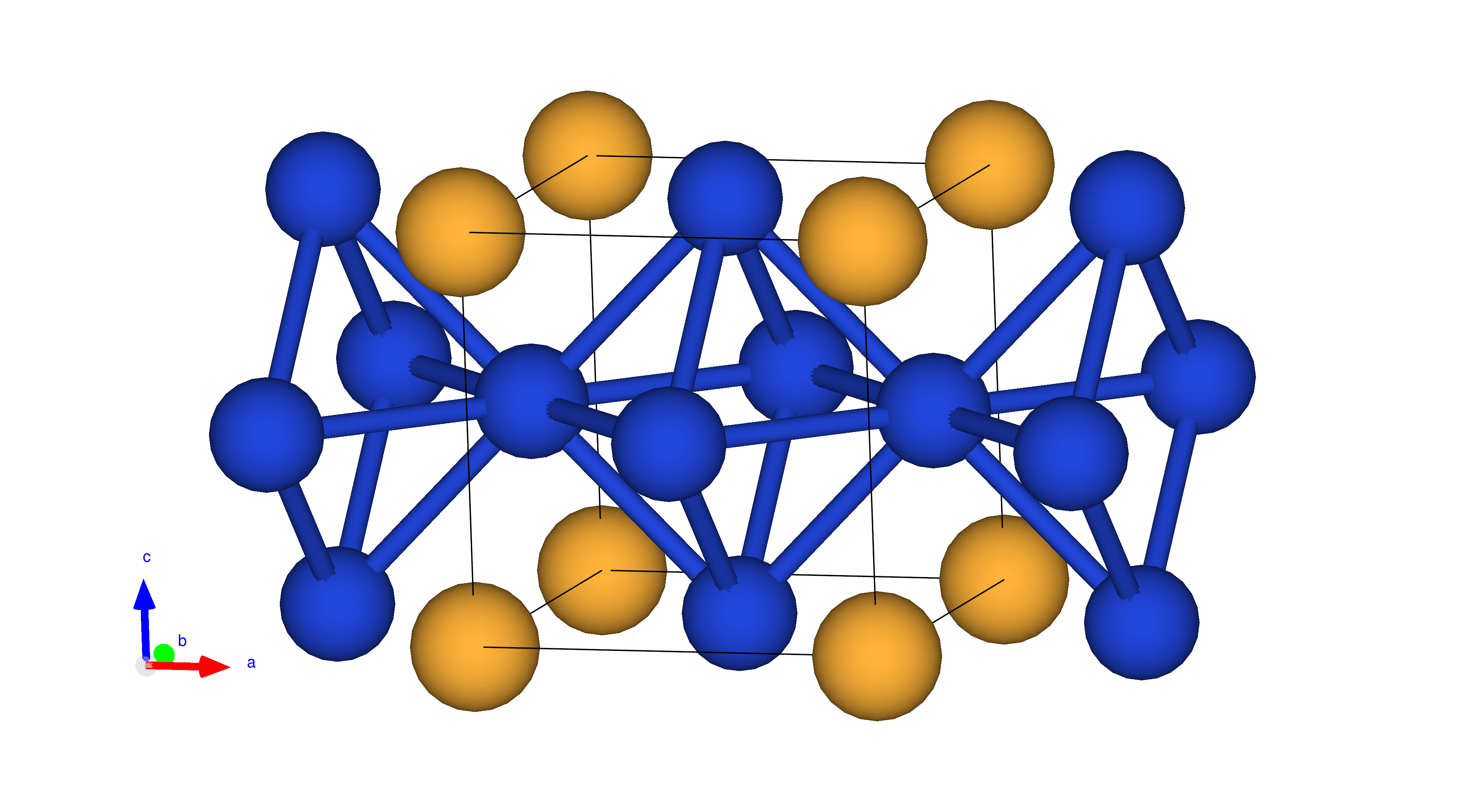}
\caption{Graphical display of the IrMn$_3$ lattice where dark blue are the Mn atoms with magnetic moments and yellow atoms are the Ir atoms without magnetic moments. One view of the lattice is through the 111 planes for the blue magnetic atoms. They form a kagome lattice. Another view is achieved by considering how each triangle shares a side with another triangle in the octahedron. This view suggests IrMn$_3$ is composed of side sharing triangles in 3 dimensions and is therefore more similar to the triangular lattice antiferromagnet than the kagome antiferromagnet.}
\label{fig:IrMn3}
\end{figure}

The use of analogs of Maxwell constraints in magnetism and the associated topological index provides a simple picture of the strength of magnetism in a system. To see this in practice, consider the question of the origin of the industrial strength of antiferromagnetism in IrMn$_3$. This material magnetically orders in a 120$^0$ coplanar state at 960 K\cite{Tomeno1999} but is composed of kagome planes with a dominant isotropic Heisenberg spin exchange\cite{Szunyogh2009} that would appear to frustrated the magnetism\cite{Chen2014}. How then can the Maxwell-like constraints advanced by this paper explain such industrial strength magnetism? The answer lies in the many different kagome planes. They all interconnect as shown in Fig. \ref{fig:IrMn3}. Indeed, the lattice is better thought of as side-sharing triangles in three dimensions instead of layers of kagome planes. With eight triangles in the unit cell, the dominant Heisenberg term in its Hamiltonian can be written
\begin{equation}
  H = \frac{J}{4}\sum_{\Delta} (\vec S_{\Delta})^2
\end{equation}
where $\vec S_{\Delta}$ is the total spin on each triangle. We therefore have a topological index $\nu = 2N_s - 3N_\Delta = (2 - 8 )N_s$ (since there are 8/3 triangles per spin). We can contrast this to say an unfrustrated antiferromagnet on the cubic lattice with three bonds in the unit cell and $\nu = 2N_s - 3N_b = (2-9)N_s = -7 N_s$.  But upon closer inspection, we should be careful with such a comparison. In IrMn$_3$, nearest neighbor Heisenberg exchange still has a sub-extensive number of zero modes  because the side sharing triangles form corner sharing octehedra. So its $T_N$ is likely lower than the unfrustrated antiferromagnet with the same $J$.  Nevertheless, this effect is subextensive so it is unlikely to be a big effect. Therefore, the energetic constraints imposed by the dominant Heisenberg exchange in IrMn$_3$ appear to be nearly as redundant as an unfrustrated antiferromagnet on a cubic lattice suggesting great industrial strength even though it is composed of kagome planes.

But not only firm magnetism may prove useful for practical applications. Fragile isostatic magnetism may prove to be easily manipulated for special design purposes.  It may play a role for magnetic systems similar to the role isostatic phonons have played in the design of metameterials. Due to the absence of redundant constraints, isostatic lattices can have particular weaknesses that can enable their manipulation. Perhaps even metamaterials designs that exploit this could carry over to magnetic materials. For example, one lattice was designed to have selective mechanical failure in Ref. \onlinecite{Paulose2015} where just a local set of bonds associated with a state of self stress broke down under a load. The magnetic analog of this could correspond to a material design with a selective spin flip of a local subset of spins achieved with a global magnetic field. In this way, isostatic magnetism could prove to have an important role in the design of magnetic materials for applications.

\begin{figure}
\includegraphics[width=0.45\columnwidth]{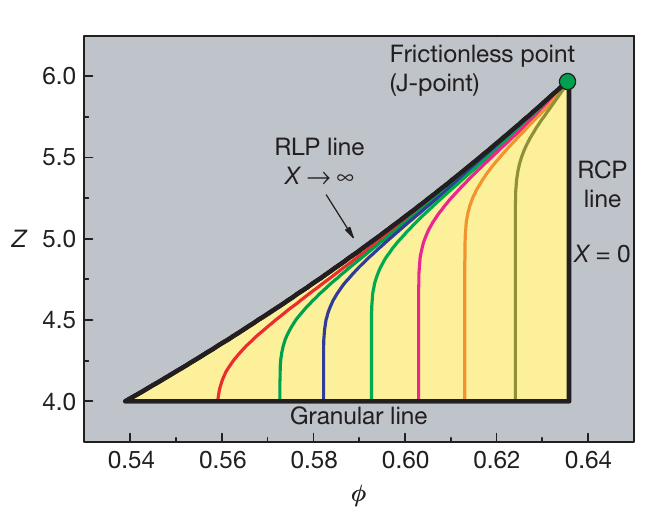}
\includegraphics[width=0.45\columnwidth]{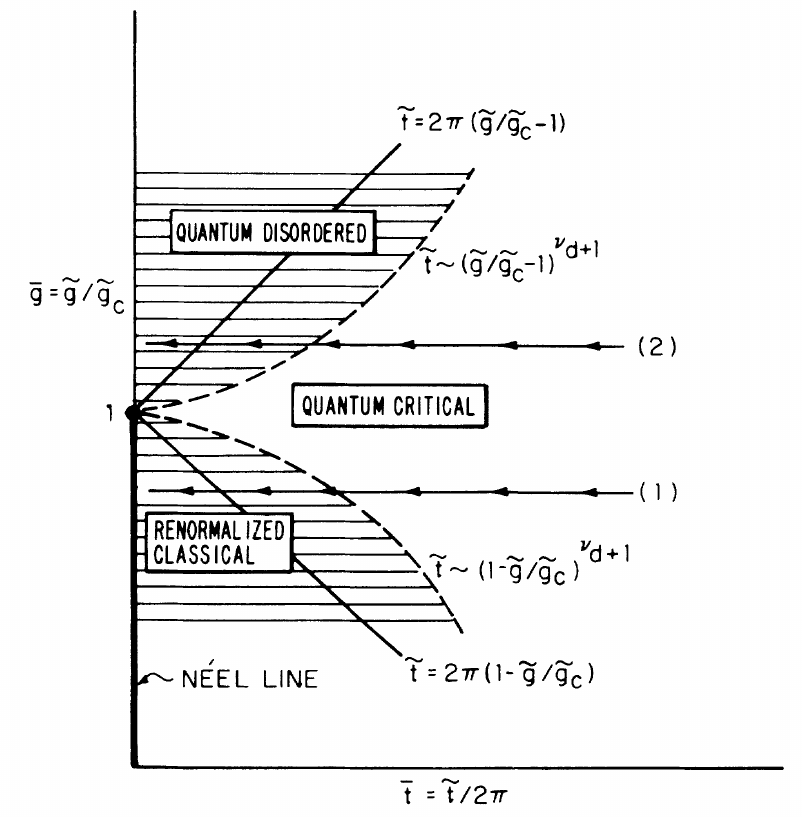}
\includegraphics[width=\columnwidth]{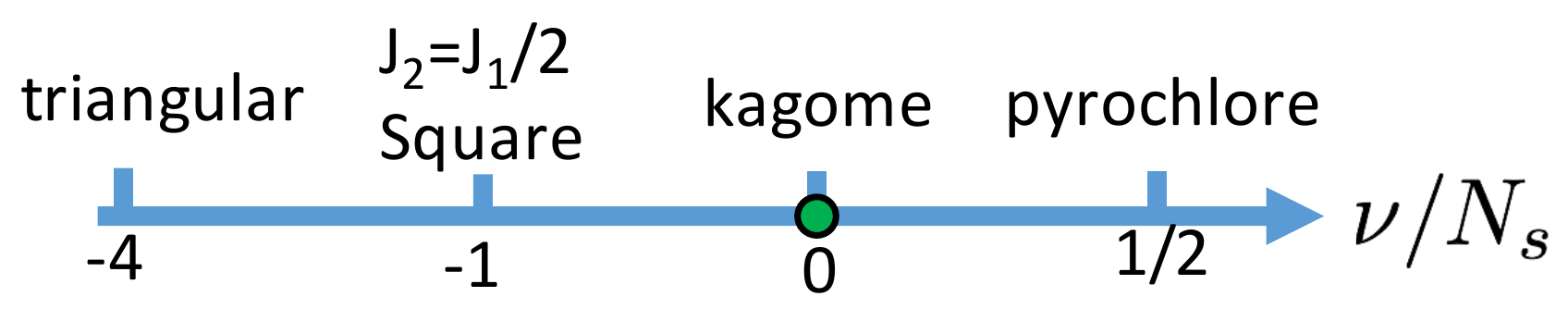}
\caption{Similarities between the jamming transition and the loss of magnetic order by either quantum effects or classical frustration. a) the jamming phase diagram taken from Ref. \cite{Song2008} with allowed jammed states for a given average coordination number $Z$ and given packing fraction $\phi$. Here the isostatic point highlighted by the green dot at the top right occurs for frictionless grains. b) the phase diagram for quantum disordering a frustrated antiferromagnet obtained in a semi-classical field theory\cite{Chakravarty1988}. The x-axis is a dimensionless temperature and the y-axis a coupling measuring the strength of quantum fluctuations. c) Classical phase diagram of the loss of magnetic order based on the results of this paper. Here again the green dot represents the isostatic point. Notice in all three cases, a special point exists in the phase diagram that sheds light on the rest of the phase diagram. It would be interesting to understand the relationship between these three points.}
\label{fig:jamming}
\end{figure}

But isostatic magnetism may also extend our theoretical understanding of magnetic materials. There is likely a separate topological property not discussed in this paper. Since the magninos are free fermions, and the ideal isostatic magnetism state is a gapped state, it must fit into our categorization of topological band theory\cite{Kitaev2009} much like the phoninos where shown to have a topological band theory by Kane and Lubensky (though they didn't call them phoninos). It is the supersymmetry established by this paper that enables physics that applies to free fermions (including Majorana fermions) to also apply to phonons and magnons. This is one of the benefits of generalizing Kane and Lubensky's theory to a broader class of systems. 

The results of this paper may also shed light on a long standing problem in the field of magnetism beyond band theory: how to melt magnetic order. The connection established by our formal developments shows explicitly that the loss of rigidity in a solid\cite{Guest2003,Hutchinson2006}, the jamming transition\cite{Cates1998,Liu1998,Song2008} and the loss of magnetic order due to frustration\ref{Ramirez1994,Moessner1998a,Moessner1998b} may all be different aspects of the same problem (see Fig. \ref{fig:jamming}).  Each of these systems are built around constraints, have a topological index $\nu$, and have isostatic order with $\nu=0$ that we curiously identify here with spontaneous SUSY breaking. So from this perspective, isostatic magnetism is at a critical point between magnetic order and a classical spin liquid. It would be interesting to understand how this critical point is related to the quantum melting of antiferromagnetic order through a quantum critical point\cite{Chakravarty1988,Haldane1988} particularly given theories of this quantum phase transition that are semi-classical in nature.  

So the discovery of an isostatic magnetic material (possibly in the family of distorted kagome antiferromagnets) could provide:
\begin{itemize}
\item practical use in the design of magnetic materials via the topological index, zero modes and self field modes
\item a novel application of topological band theory of fermions to magnons
\item insight into the mechanism of quantum spin liquids and other exotic phases of magnetism. 
\end{itemize}
Remarkably, all of this was achieved through the use of supersymmetry in condensed matter physics to draw formal connections between two seemingly different physical problems. 

{\bf Acknowledgement} We would like to thank Tom Lubensky, Charlie Kane, Vincenzo Vitelli, Tom Hartman and Allan MacDonald for useful discussions. 

%

\end{document}